\documentclass[showpacs,preprintnumbers,amsmath,amssymb,prb,superscriptaddress,longbibliography]{revtex4-2}

\usepackage{graphicx}
\usepackage{dcolumn}
\usepackage{bm}

\begin{document}

\title{Field-induced transitions from incommensurate to commensurate phases in helical antiferromagnets}

\author{P.\ T.\ Bolokhova}
\email{polina.bolokhova@gmail.com}
\affiliation{National Research University Higher School of Economics, St.\ Petersburg 190121, Russia}
\affiliation{Petersburg Nuclear Physics Institute named by B.P.\ Konstantinov of National Research Center ‘‘Kurchatov Institute’’, Gatchina 188300, Russia}

\author{A.\ V.\ Syromyatnikov}
\email{asyromyatnikov@yandex.ru}
\affiliation{Petersburg Nuclear Physics Institute named by B.P.\ Konstantinov of National Research Center ‘‘Kurchatov Institute’’, Gatchina 188300, Russia}

\date{\today}

\begin{abstract}

Heisenberg antiferromagnet with an easy-plane anisotropy is discussed in which a magnetic spiral is induced by Dzyaloshinskii-Moriya interaction and/or frustration of the exchange coupling. The distortion of the spiral by small in-plane magnetic field is described analytically. It is found that the field can gradually change the vector of the magnetic structure ${\bf k}_0$ and can produce transitions between phases with incommensurate and commensurate magnetic orderings when ${\bf k}_0$ is close to ${\bf g}/n$, where ${\bf g}$ is a reciprocal lattice vector and $n$ is integer. Analytical expressions for critical fields are derived for $n=2$, 3, and 4. Application of the theory to the triangular-lattice compound $\rm RbFe(MoO_4)_2$ is discussed alongside its potential applicability to other materials. As a by-product of the main consideration, model parameters are found which describe more accurately the full set of available experimental data suggested before for $\rm RbFe(MoO_4)_2$.

\end{abstract}


\maketitle

\section{Introduction}

Noncollinear magnetic systems are attracting considerable interest now because they may host topologically nontrivial spin orderings including different types of magnetic solitons \cite{bogdanov2020,fert2017} and multiferroic states induced by the noncollinear magnetic order \cite{spaldin2019,Tokura2014,Dong02112015,Vopson04072015}. The ability to easily control properties of such materials using electric current and/or external fields makes promising their technological application in spintronics, memory devices, and non-traditional computing. \cite{fert2017,D0MH01603A,spaldin2019,Vopson04072015} This makes it currently important to study noncollinear magnets in external magnetic field and field-induced phase transitions in them.

It is well known that magnetic systems can exhibit magnetically ordered states that are either commensurate or incommensurate with respect to the underlying crystal lattice. Phase transitions between incommensurate and commensurate phases (hereafter referred to as IC transitions) are also observed in some systems upon varying temperature and/or external fields (see Ref.~\cite{bak} for a survey of such transitions in very different physical systems). \cite{dzya64, Izyumov1984, Nagamiya1962, maslov} The underlying mechanism driving such transitions is widely recognized as a competition between ''built-in'' spatial periods characteristic to different terms in the Hamiltonian.\cite{bak}  

A remarkable theoretical consideration of the IC transition in a magnetic system was performed by Dzyaloshinskii in Ref.~\cite{dzya64}. A layered easy-plane ferromagnet was discussed there in which a small Dzyaloshinskii-Moria interaction (DMI) acting between spins from neighboring layers produces a long-period magnetic spiral. It was known that the magnetic field $\bf h$ applied in the easy plane deforms the spiral so that the resultant periodic magnetic structure looks like a soliton lattice which is described by a set of harmonics of the zero-field vector of the magnetic structure ${\bf k}_0$ (see, e.g., Ref.~\cite{Nagamiya}). A second-order IC transition was predicted in Ref.~\cite{dzya64} at a critical field $h_c$ at which the period of the soliton lattice becomes infinite and the system comes into the forced ferromagnetic state. Such transition was subsequently observed experimentally, e.g., in spiral ferromagnet $\rm CrNb_3S_6$ (see, e.g., Ref.\cite{sasha} and references therein). 

A similar theory of field-induced IC transitions based on an analysis of the classical ground state energy in the continuum approximation was suggested in Ref.~\cite{maslov} for easy-plane antiferromagnets. It was assumed in that theory that a small DMI produces a small deviation of ${\bf k}_0$ from the vector describing the collinear N\'eel state (which is equal to ${\bf g}/2$, where $\bf g$ is a reciprocal lattice vector). Similar to ferromagnetic systems, a second-order IC transition was predicted in Ref.~\cite{maslov} upon the field increasing from the soliton state to the commensurate canted antiferromagnetic phase described by ${\bf g}/2$. However many experimental results are available now which evidence a first-order character of such transitions in antiferromagnetic spiral magnets (see, e.g., Refs.~\cite{bacuge,exp1,exp2,exp3,exp4,rbfemoo3,boratei}). It was shown in Ref.~\cite{martynov} that higher order derivatives in equations for spins angles which were ignored in Ref.~\cite{maslov} make this transition discontinuous.

In the present paper, we carry out quite a general theoretical consideration of IC transitions in easy-plane helical antiferromagnets induced by in-plane magnetic field which is much smaller than the saturation field $h_s$. We assume that the spiral order appears in the zero field due to the frustration of exchange spin couplings and/or small DMI and that ${\bf k}_0$ is close to ${\bf g}/n$, where $n$ is integer. 
\footnote{Of course, we assume that ${\bf g}/n$ is not equal to a reciprocal lattice vector.}
By comparing energies of spiral and commensurate states found in the fourth order in $h/h_s\ll1$, we derive general expressions for the critical field $h_c$ of the first order transition from the (soliton) state with distorted spiral to commensurate states whose ordering is described by the vector ${\bf g}/n$. The closeness of ${\bf k}_0$ to ${\bf g}/n$ provides the smallness of $h_c/h_s$ which is required for the validity of the present consideration. We discuss in detail cases of $n=2$, 3, and 4. 
\footnote{In the case of $n=4$, it is assumed also that ${\bf g}/2$ is not equal to a reciprocal lattice vector.}
We describe analytically magnetic orderings in the incommensurate and commensurate phases thus refining results of Refs.\cite{dzya64, maslov, Izyumov1984, Nagamiya1962}. We also show that the field can gradually change the vector of the magnetic structure ${\bf k}_0$ in the incommensurate state as it is observed in some experiments (see, e.g., Refs.~\cite{rbfemoo4,mitamura}).

Notice that the case of $n=2$ was considered in Refs.~\cite{maslov} in relation with corresponding experiments in $\rm Ba_2CuGe_2O_7$. However it turns out that this material shows a more complicated behavior near $h_c$ due to the smallness of the easy-plane anisotropy (as a result, spins come out of the easy plane and an additional phase arises near $h_c$). \cite{papa,bacuge} The application of the developed theory at $n=2$ is more straightforward in the case of NdFe$_{3}$(BO$_{3}$)$_{4}$ as it is demonstrated in the related paper \cite{boratei}. As $n=3$ is relevant to triangular-lattice compounds, we apply our theory below for description of experimental results in the corresponding material $\rm RbFe(MoO_4)_2$. We are not aware of experimental findings corresponding to the discussed transition at $n=4$.

The rest of the present paper is organized as follows. In Sec.~\ref{gen}, we present the model Hamiltonian and consider it in the absence of the magnetic field. In Sec.~\ref{sf}, we apply a small magnetic field in the easy plane and discuss the general consequences it induces. Sec~\ref{higher} is dedicated to more detailed calculations of the ground state energy and magnetic ordering specific to each phase. In Sec.~\ref{transit}, we compare the ground state energies of incommensurate and commensurate phases, derive critical fields, consider quantum and themral corrections to critical fields and explain details of IC phase transitions. In Sec.~\ref{app}, we apply our results to $\rm RbFe(MoO_4)_2$, the triangular-lattice antiferromagnet in which the first-order IC phase transition was observed in Refs. \cite{rbfemoo1,rbfemoo3}. As a by-product of the main consideration, we scrutinize the model suggested before for $\rm RbFe(MoO_4)_2$ and find model parameters which describe more accurately the full set of available experimental data. In Sec.~\ref{conc}, the summary of our findings is given. In Appendix~\ref{est}, we consider a very simple particular model which supports some claims made in the main text. In Appendix~\ref{clim}, we show by the example of a simple model that the transition to the IC phase at $n=3$ is governed by a sine-Gordon equation in the continuous limit in the leading order in $h$ (that implies a first-order nature of this transition according to conclusions of Ref.~\cite{martynov} due to higher order in $h$ terms containing higher order derivatives). In Appendix~\ref{clim_arb}, we carry out a general consideration of the simple model and come to the sine-Gordon equation for arbitrary $n\ge2$.

\section{General consideration. Zero field.}
\label{gen}

We discuss a spiral magnet described by the Hamiltonian 
\begin{equation}
 \label{ham0}
 \mathcal{H} = 
\frac12 \sum_{j,{\bf b}} 
\Bigl(
J_{\bf b} \left({\bf S}_{{\bf r}_j} {\bf S}_{{\bf r}_j+{\bf b}}\right)
+
{\bf D}_{\bf b} \left[{\bf S}_{{\bf r}_j} \times {\bf S}_{{\bf r}_j+{\bf b}}\right]
\Bigr)
+
A\sum_{j} \left(S^y_{{\bf r}_j}\right)^2,
\end{equation}
where $A>0$ is the easy-plane anisotropy value which confines spins to lie within $xz$ plane and we imply that spin ${\bf S}_{{\bf r}_j}$ at site ${\bf r}_j$ is coupled with its neighbor at site ${\bf r}_j+{\bf b}$ with exchange constant $J_{\bf b}$ and Dzyaloshinky-Moria interaction (DMI) with vector ${\bf D}_{\bf b}=-{\bf D}_{-\bf b}$. We will assume for simplicity that ${\bf D}_{\bf b}$ is directed along $y$ axis. The spin helix can arise in this model due to a frustration in exchange constants $J_{\bf b}$ or a competition between the exchange and the DMI couplings (or as a result of action of both of these mechanisms). 

Spin components in the laboratory coordinate frame $(x,y,z)$ are related as follows with spin components in the local frame $(x',y',z')$ in which $z'$ axis is directed along the average local magnetization (whose value is usually reduced by quantum and thermal fluctuations as a result of taking into account $1/S$ corrections in the developed spin-wave theory, where $S$ is the spin value):
\begin{eqnarray}
  \label{transform}
	S^{x}_{{\bf r}_j} &=& S^{x^\prime}_{{\bf r}_j} \cos({\bf k}_0{\bf r}_j+\phi) + S^{z^\prime}_{{\bf r}_j} \sin({\bf k}_0{{\bf r}_j}+\phi),\nonumber\\
  S^{y}_{{\bf r}_j} &=& S^{y^\prime}_{{\bf r}_j} ,\\
	S^{z}_{{\bf r}_j} &=& -S^{x^\prime}_{{\bf r}_j} \sin({\bf k}_0{\bf r}_j+\phi) + S^{z^\prime}_{{\bf r}_j} \cos({\bf k}_0{\bf r}_j+\phi),\nonumber
\end{eqnarray}
where $\phi$ is a constant and ${\bf k}_0$ is the vector of the magnetic structure. Then, the magnetic order is described as
\begin{eqnarray}
  \label{oder0}
	S^{x}_{{\bf r}_j} &=& S \sin({\bf k}_0{{\bf r}_j}+\phi),\nonumber\\
	S^{y}_{{\bf r}_j} &=& 0,\\
	S^{z}_{{\bf r}_j} &=& S \cos({\bf k}_0{{\bf r}_j}+\phi).\nonumber
\end{eqnarray}

Substituting Eqs.~\eqref{transform} to Eq.~\eqref{ham0} and taking the Fourier transform, one obtains for the Hamiltonian
\begin{equation}
 \label{ham}
 \mathcal{H} = 
\sum_{\bf k} 
\left[
\left(J_{\bf k}+A\right) S^{y'}_{\bf k} S^{y'}_{-\bf k} 
+
\tilde J_{\bf k} \left( S^{x'}_{\bf k} S^{x'}_{-\bf k} + S^{z'}_{\bf k} S^{z'}_{-\bf k} \right)
-2i
J'_{\bf k}  S^{x'}_{\bf k} S^{z'}_{-\bf k}
\right],
\end{equation}
where
\begin{eqnarray}
  \label{js}
J_{\bf k} &=& \frac12 \sum_{\bf b} J_{\bf b} \cos({\bf kb}),\nonumber\\
\tilde J_{\bf k} &=& \frac12 \sum_{\bf b} J_{\bf b} \cos({\bf k_0}{\bf b}) \cos({\bf kb}) + \frac12 \sum_{\bf b} D_{\bf b} \sin({\bf k_0}{\bf b}) \cos({\bf kb}) = \frac12 \sum_{\bf b} \sqrt{J_{\bf b}^2 + D_{\bf b}^2} \cos({\bf k'_0} {\bf b}) \cos (\bf kb),\\
J'_{\bf k} &=& \frac12 \sum_{\bf b} J_{\bf b} \sin({\bf k_0}{\bf b}) \sin({\bf kb}) - \frac12 \sum_{\bf b} D_{\bf b} \cos({\bf k_0}{\bf b}) \sin({\bf kb}) = \frac12 \sum_{\bf b} \sqrt{J_{\bf b}^2 + D_{\bf b}^2} \sin({\bf k'_0} {\bf b}) \sin (\bf kb),\nonumber\\ 
{\bf k'_0} {\bf b} &=& {\bf k_0} {\bf b} - \arcsin \left({\frac{D_{\bf b}}{\sqrt{J_{\bf b}^2 + D_{\bf b}^2}}}\right).\nonumber
\end{eqnarray}
It is seen from Eq.~\eqref{ham} that the classical ground state energy has the form
\begin{equation}
	\label{gse0}
	\frac{{\cal E}_{h=0}}{N} = S^2\tilde J_{\bf 0},
\end{equation}
where $N$ is the number of spins in the crystal (we consider for simplicity a Bravais lattice with one spin in the unit cell). Then, minimization of $\tilde J_{\bf 0}$ given by Eq.~\eqref{js} defines the vector of the magnetic structure ${\bf k}_0$ at zero field which is in general incommensurate.

In further consideration, it is convenient to adopt the Holstein-Primakoff spin representation
\begin{eqnarray}
  \label{hp}
	S^{x^{\prime}}_{{\bf r}_j}
	&\approx&
	\sqrt{\frac{S}{2}}\left(a^\dagger_{{\bf r}_j}+a^{}_{{\bf r}_j} - a^\dagger_{{\bf r}_j}\frac{a^\dagger_{{\bf r}_j}+a^{}_{{\bf r}_j}}{4S}a^{}_{{\bf r}_j}\right), \nonumber \\
  S^{y^{\prime}}_{{\bf r}_j}
	&\approx& 
	i\sqrt{\frac{S}{2}}\left(a^\dagger_{{\bf r}_j}-a^{}_{{\bf r}_j} - a^\dagger_{{\bf r}_j}\frac{a^\dagger_{{\bf r}_j}-a^{}_{{\bf r}_j}}{4S}a^{}_{{\bf r}_j}\right),\\
	S^{z^{\prime}}_{{\bf r}_j} &=& S-a^\dagger_{{\bf r}_j} a^{}_{{\bf r}_j}.  \nonumber
\end{eqnarray}
After substitution of Eqs.~\eqref{hp} to Eq.~\eqref{ham}, one obtains that the Hamiltonian has the form
\begin{equation}
\label{hamhp}
	{\cal H} = {\cal E}_0 +\sum_{i=2}^\infty {\cal H}_i,
\end{equation}
where ${\cal E}_0$ is the classical ground state energy and ${\cal H}_i$ are terms containing products of $i$ Bose-operators. In particular, one has
\begin{eqnarray}
\label{h2}
	\mathcal{H}_2 &=& 
	\sum_{\bf k} 
	\left[ 
	a^\dagger_{\bf k} a^{}_{\bf k} 
	S \left( J_{\bf k} + \tilde J_{\bf k} - 2 \tilde J_{\bf 0} + \tilde A \right) 
	+
	\left( a^{}_{\bf k} a^{}_{-\bf k}  + a^\dagger_{\bf k} a^\dagger_{-\bf k}  \right)
	\frac S2 \left( \tilde J_{\bf k} - J_{\bf k} - \tilde A \right)
	\right],\\
\label{h3}
	\mathcal{H}_3 &=& 
	i\sqrt{\frac{2S}{N}} \sum_{{\bf k}_1+{\bf k}_2+{\bf k}_3={\bf 0}}
	J'_{{\bf k}_1} a^\dagger_{-\bf k_2} \left(a^\dagger_{-\bf k_1} + a^{}_{\bf k_1}\right) a^{}_{\bf k_3},
\end{eqnarray}
where 
\begin{equation}
	\label{at}
	\tilde A=A\left(1-\frac{1}{2S}\right).
\end{equation}
Notice that the value of the easy-plane anisotropy should enter in all expressions for observable quantities as $\tilde A$ (giving zero at $S=1/2$) because the last term in Eq.~\eqref{ham0} turns into a constant at $S=1/2$. \cite{chubrev} One obtains from Eq.~\eqref{h2} for the magnon spectrum in the linear spin-wave approximation
\begin{equation}
	\label{spec}
	\epsilon_{\bf k} = 2S\sqrt{\left( \tilde J_{\bf k} - \tilde J_{\bf 0} \right)
	\left( J_{\bf k} - \tilde J_{\bf 0} + \tilde A\right)}.
\end{equation}

\section{Small field in the easy plane. Leading field corrections.}
\label{sf}

Let us apply a small field ${\bf h}$ in the easy plane. Due to the arbitrariness of $\phi$ in Eq.~\eqref{transform}, we can assume that the field is parallel to $x$ axis and the Zeeman term in the Hamiltonian has the form
\begin{equation}
	\label{zeeman}
	\mathcal{H}_Z = -h\sum_j S_j^x.
\end{equation}
It is seen from Eqs.~\eqref{transform} and \eqref{hp} that $\mathcal{H}_Z$ contributes into terms in the Hamiltonian which are linear and quadratic in Bose operators. These terms acquire the form
\begin{eqnarray}
\label{hz1}
	\mathcal{H}_{Z1} &=& -h \sqrt{\frac{SN}{8}}
	\left[ 
	e^{i\phi}\left(a^\dagger_{\bf k_0} + a^{}_{-\bf k_0}\right) 
	+
	e^{-i\phi}\left(a^\dagger_{-\bf k_0} + a^{}_{\bf k_0}\right)
	\right],\\
\label{hz2}
	\mathcal{H}_{Z2} &=& -h \frac i2 \sum_{\bf k}
	\left[ 
	e^{i\phi}a^\dagger_{\bf k+k_0} a^{}_{\bf k} 
	-
	e^{-i\phi}a^\dagger_{\bf k-k_0} a^{}_{\bf k}
	\right],
\end{eqnarray}
respectively. The appearance of linear terms $\mathcal{H}_{Z1}$ signifies that the spin ordering described by Eqs.~\eqref{oder0} is distorted by the finite field. To dispose of linear terms \eqref{hz1} in the Hamiltonian and to find field corrections to observables, we perform the following shift:
\begin{eqnarray}
\label{shift}
a^{}_{\bf k_0} &\mapsto& a^{}_{\bf k_0} + \rho_{+}\sqrt{NS},\nonumber\\
a^{}_{-\bf k_0} &\mapsto& a^{}_{-\bf k_0} + \rho_{-}\sqrt{NS},\\
a^\dagger_{\bf k_0} &\mapsto& a^\dagger_{\bf k_0} + \rho^*_{+}\sqrt{NS},\nonumber\\
a^\dagger_{-\bf k_0} &\mapsto& a^\dagger_{-\bf k_0} + \rho^*_{-}\sqrt{NS},\nonumber
\end{eqnarray}
where $\rho_\pm$ are complex parameters. Linear terms in $a^{}_{\pm\bf k_0}$ and $a^\dagger_{\pm\bf k_0}$ arise from ${\cal H}_2$ (Eq.~\eqref{h2}) after performing shift \eqref{shift} which cancel $\mathcal{H}_{Z1}$ (Eq.~\eqref{hz1}) if
\begin{eqnarray}
\label{rhog}
\rho_{-} &=& \rho^*_{+} = \rho e^{-i\phi},\\
\label{rho1}
\rho &=& h \frac{1}{4\sqrt2S}\frac{1}{\tilde J_{\bf k_0} - \tilde J_{\bf 0}}.
\end{eqnarray}
After substitution of Eqs.~\eqref{shift} into Eqs.~\eqref{h2} and \eqref{hz1}, one finds the correction to the ground state energy \eqref{gse0} in the leading order in $h$
\begin{equation}
	\label{gse}
	\frac{{\cal E}_h}{N} = S^2 \tilde J_{\bf 0} - \frac{hS\rho}{\sqrt2}
	= S^2 \tilde J_{\bf 0} - \frac{h^2}{8\left(\tilde J_{\bf k_0} - \tilde J_{\bf 0}\right)}.
\end{equation}

The physical consequence of shift \eqref{shift} is the additional rotation of each spin from its position determined by Eq.~\eqref{oder0} with the formation of the distorted magnetic ordering. The latter can be easily obtained in the first order in $h$ using Eqs.~\eqref{transform}, \eqref{hp}, and \eqref{shift}, the result being
\begin{eqnarray}
  \label{oderh}
	S^{x}_{{\bf r}_j} &=& 
	S \left[ 
	2\sqrt2\rho\cos^2({\bf k}_0{\bf r}_j+\phi) 
	+ 
	\sin({\bf k}_0{{\bf r}_j}+\phi) 
	\right],\\
	S^{z}_{{\bf r}_j} &=& 
	S \left[ 
	-2\sqrt2\rho\sin({\bf k}_0{\bf r}_j+\phi)\cos({\bf k}_0{\bf r}_j+\phi) 
	+ 
	\cos({\bf k}_0{{\bf r}_j}+\phi) 
	\right],\nonumber
\end{eqnarray}
where $\rho$ is given by Eq.~\eqref{rho1}. It can be checked straightforwardly that $(S^{x}_{{\bf r}_j})^2 + (S^{z}_{{\bf r}_j})^2 =S^2(1+{\cal O}(h^2))$. Interestingly, Eqs.~\eqref{oderh} are equivalent within the first order in $h$ to more compact expressions (cf.\ Eqs.~\eqref{oder0}):
\begin{eqnarray}
  \label{order2}
	S^{x}_{{\bf r}_j} &=& 
	S \sin\left( ({\bf k}_0{\bf r}_j+\phi) + 2\sqrt2 \rho \cos({\bf k}_0{\bf r}_j+\phi)
	\right),\\
	S^{z}_{{\bf r}_j} &=& 
	S \cos\left( ({\bf k}_0{\bf r}_j+\phi) + 2\sqrt2 \rho \cos({\bf k}_0{\bf r}_j+\phi)
	\right).\nonumber
\end{eqnarray}
Similar expressions were derived before \cite{Nagamiya} in the first order in $h$.

It is easy to realize from Eqs.~\eqref{rho1} and \eqref{gse} that magnetic field can smoothly change the vector of the magnetic structure (see Appendix~\ref{est} for more detail). It happens when linear terms are finite in the expansion of $\tilde J_{\bf k}$ near ${\bf k}={\bf k}_0$. In this case, Eq.~\eqref{gse} is minimized at ${\bf k}_0={\bf k}^{(0)}_0+\delta{\bf k}_0$, where $\delta k_0\propto h^2$ and ${\bf k}^{(0)}_0$ is the vector of the magnetic structure at $h=0$ (we take into account here that $\tilde J_{\bf 0}$ is quadratic in $\delta{\bf k}_0$ because $\tilde J_{\bf 0}$ is minimized at ${\bf k}_0={\bf k}_0^{(0)}$ at $h=0$). This correction to ${\bf k}_0$ is accompanied with terms in the energy of the order of $h^4$ which we discuss below.

In the opposite case, when $\tilde J_{\bf k}$ is quadratic near ${\bf k}={\bf k}_0$, the vector of the magnetic structure is not effected by the magnetic field in the considered order in $h$. However we demonstrate below that further terms in powers of $h$ in the ground state energy can lead to drastic changes of magnetic orderings: higher-order terms turn out to be smaller for commensurate ${\bf k}_0$ so that transitions can happen at large enough $h$ from incommensurate to commensurate phases. To find higher order field corrections to the ground state energy, one has to consider terms in the Hamiltonian given by Eqs.~\eqref{h3}, \eqref{hz2} as well as ${\cal H}_i$ in Eq.~\eqref{hamhp} with $i>3$. These terms produce contributions linear in $a^{}_{\pm p{\bf k}_0}$ and $a^\dagger_{\pm p{\bf k}_0}$ after shift \eqref{shift}, where $p\ge2$ is integer. Then, further consideration differs drastically for incommensurate ${\bf k}_0$ and commensurate ${\bf k}_0={\bf g}/n$, where ${\bf g}$ is a reciprocal lattice vector, because, for instance, $a^{}_{\pm n{\bf k}_0}$ is equivalent in the latter case to $a^{}_{\bf 0}$, $a^{}_{\pm (n-1){\bf k}_0}$ is equivalent to $a^{}_{\mp {\bf k}_0}$, etc. We consider all these cases in the next section for $n=2$, 3, and 4.

\section{Higher order field corrections}
\label{higher}

\subsection{Incommensurate ${\bf k}_0$}

If ${\bf k}_0$ is incommensurate, Eqs.~\eqref{h3} and \eqref{hz2} produce terms in the Hamiltonian proportional to $a^{}_{\pm2{\bf k}_0}$ and $a^\dagger_{\pm2{\bf k}_0}$ which are the order of $h^2$. One has to make the following shift to cancel them (cf.\ Eq.~\eqref{shift}):
\begin{eqnarray}
\label{shift2}
a^{}_{2\bf k_0} &\mapsto& a^{}_{2\bf k_0} + \gamma_{+}\sqrt{NS},\nonumber\\
a^{}_{-2\bf k_0} &\mapsto& a^{}_{-2\bf k_0} + \gamma_{-}\sqrt{NS},\\
a^\dagger_{2\bf k_0} &\mapsto& a^\dagger_{2\bf k_0} + \gamma^*_{+}\sqrt{NS},\nonumber\\
a^\dagger_{-2\bf k_0} &\mapsto& a^\dagger_{-2\bf k_0} + \gamma^*_{-}\sqrt{NS}.\nonumber
\end{eqnarray}
Straightforward calculations similar to those carried out above for $\rho_\pm$ give
\begin{eqnarray}
\gamma_{-} &=& \gamma^*_{+} = \gamma ie^{-2i\phi},\\
\label{gamma}
\gamma &=& \rho^2 
\frac{2\tilde J_{\bf 0}-2\tilde J_{\bf k_0}+2 J'_{\bf k_0} - J'_{2 \bf k_0}}{\sqrt2\left(\tilde J_{2\bf k_0} - \tilde J_{\bf 0}\right)}. 
\end{eqnarray}
The corresponding correction to the ground state energy ${\cal E}_h^{(i)}$ is of the order of $h^4$ so that ${\cal E}_h^{(i)}$ has the form \eqref{gse} in the third order in $h$.

To find the remaining contributions of the order of $h^4$ to ${\cal E}_h^{(i)}$, one has to cancel terms in the Hamiltonian proportional to $a_{\pm p {\bf k}_0}$ and $a_{\pm p {\bf k}_0}^{\dagger}$, where $p = 3, 4$, which are of the orders of $h^3$ and $h^4$, by making shifts in $a_{\pm 3 {\bf k}_0}$, $a_{\pm 3 {\bf k}_0}^{\dagger}$, $a_{\pm 4 {\bf k}_0}$, and $a_{\pm 4 {\bf k}_0}^{\dagger}$ similar to Eqs.~\eqref{shift} and \eqref{shift2}. Besides, one has to take into account corrections to $\rho$ and $\gamma$ of the orders of $h^3$ and $h^4$ and to calculate the contribution to ${\cal E}_h^{(i)}$ stemming from the field-induced correction $\delta{\bf k}_0$ to ${\bf k}_0^{(0)}$ discussed above. To obtain the latter quantity, we minimize Eq.~\eqref{gse} with respect to $\delta{\bf k}_0$ assuming for simplicity that $\tilde J_{\bf 0}$ and $\tilde J_{\bf k_0}$ have the form 
\begin{eqnarray}
\label{dj0}
	\tilde J_{\bf 0} &\approx& \tilde J_{\bf 0}^{(0)} + \alpha_x\delta k_x^2 + \alpha_y\delta k_y^2 + \alpha_z\delta k_z^2,\\
\label{dtj0}
	\tilde J_{\bf k_0} &\approx& \tilde J_{\bf k_0}^{(0)} + \beta_x\delta k_x + \beta_y\delta k_y + \beta_z\delta k_z,
\end{eqnarray}
where $\delta{\bf k}_0=(\delta k_x,\delta k_y,\delta k_z)$, $\tilde J_{\bf 0}^{(0)}$ and $\tilde J_{\bf k_0}^{(0)}$ are values of $\tilde J_{\bf 0}$ and $\tilde J_{\bf k_0}$ at $h=0$, respectively, and $\alpha_{x,y,z}>0$ and $\beta_{x,y,z}$ are constants. As a result, one obtains after simple but tedious calculations
\begin{eqnarray}
	\label{gsei4}
	\frac{{\cal E}_h^{(i)}}{N} &=& 
	S^2 \tilde J_{\bf 0} 
-\frac{hS\rho}{\sqrt2}\\
    &&{} + 
		h^4
		\left(
		\frac{\left(J_{\bf 0} - J_{{\bf k}_0} + \tilde J_{\bf 0} - 2 \tilde J_{{\bf k}_0} + \tilde J_{2{\bf k}_0}\right)\left(J_{\bf 0} - J_{{\bf k}_0} + 2 \tilde J_{\bf 0} - 2 \tilde J_{{\bf k}_0}\right)}{128 S^2 \left(\tilde J_{\bf 0} - \tilde J_{{\bf k}_0}\right)^4\left(\tilde J_{\bf 0} - \tilde J_{2{\bf k}_0}\right)}
		- 
		\frac{\alpha_y\alpha_z\beta_x^2+\alpha_x\alpha_z\beta_y^2+\alpha_y\alpha_x\beta_z^2}{256 S^2\alpha_x\alpha_y\alpha_z \left(\tilde J_{\bf 0} - \tilde J_{{\bf k}_0}\right)^4}
		\right)
	 + {\cal O}(h^5).\nonumber
\end{eqnarray}
The first term in brackets in Eq.~\eqref{gsei4} originates from (i) shifts \eqref{shift} and \eqref{shift2} in which terms up to the fourth order in $h$ are taken into account in $\rho$ and $\gamma$, and (ii) shifts in $a_{\pm 3 {\bf k}_0}$, $a_{\pm 3 {\bf k}_0}^{\dagger}$, $a_{\pm 4 {\bf k}_0}$, and $a_{\pm 4 {\bf k}_0}^{\dagger}$ discussed above. The second term in brackets in Eq.~\eqref{gsei4} is due to the field-induced correction to ${\bf k}_0^{(0)}$ stemming from Eqs.~\eqref{dj0} and \eqref{dtj0} in calculation of which only linear in $h$ corrections to $\rho$ have to be taken into account.

The magnetic ordering reads in the second order in $h$ as
\begin{eqnarray}
  \label{oderhi}
	S^{x}_{{\bf r}_j} &=& 
	S \left[ 
	2\sqrt2 \cos({\bf k}_0{\bf r}_j+\phi) 
	\left[
	\rho \cos({\bf k}_0{\bf r}_j+\phi) + \gamma \sin(2{\bf k}_0{\bf r}_j+2\phi)
	\right]
	+ 
	\sin({\bf k}_0{{\bf r}_j}+\phi) \left(1-4\rho^2\cos^2({\bf k}_0{\bf r}_j+\phi) \right)
	\right],\\
	S^{z}_{{\bf r}_j} &=& 
	S \left[ 
	-2\sqrt2 \sin({\bf k}_0{\bf r}_j+\phi)
	\left[
	\rho \cos({\bf k}_0{\bf r}_j+\phi) + \gamma \sin(2{\bf k}_0{\bf r}_j+2\phi)
	\right]
	+ 
	\cos({\bf k}_0{{\bf r}_j}+\phi) \left(1-4\rho^2\cos^2({\bf k}_0{\bf r}_j+\phi) \right)
	\right],\nonumber
\end{eqnarray}
where $\rho$ and $\gamma$ are given by Eqs.~\eqref{rho1} and \eqref{gamma}, respectively, and $\phi$ remains arbitrary because the ground state energy \eqref{gsei4} does not depend on it. It can be checked straightforwardly that $(S^{x}_{{\bf r}_j})^2 + (S^{z}_{{\bf r}_j})^2 =S^2(1+{\cal O}(h^3))$. Similar to Eqs.~\eqref{order2}, Eqs.~\eqref{oderhi} are equivalent within the second order in $h$ to the following more compact expressions (cf.\ Eqs.~\eqref{oder0} and \eqref{order2}):
\begin{eqnarray}
  \label{order22}
	S^{x}_{{\bf r}_j} &=& 
	S \sin\left( ({\bf k}_0{\bf r}_j+\phi) + 2\sqrt2 \rho \cos({\bf k}_0{\bf r}_j+\phi)
	+
	2\sqrt2 \gamma \sin2({\bf k}_0{\bf r}_j+\phi)
	\right),\\
	S^{z}_{{\bf r}_j} &=& 
	S \cos\left( ({\bf k}_0{\bf r}_j+\phi) + 2\sqrt2 \rho \cos({\bf k}_0{\bf r}_j+\phi)
	+
	2\sqrt2 \gamma \sin2({\bf k}_0{\bf r}_j+\phi)
	\right).\nonumber
\end{eqnarray}

\subsection{${\bf k}_0={\bf g}/2$}

Let us consider a commensurate state in which the vector of the magnetic ordering ${\bf k}_0={\bf g}/2$, where ${\bf g}$ is a reciprocal lattice vector. In this case, $a_{p {\bf k}_0}$ ($a^{\dagger}_{p {\bf k}_0}$) is equivalent to either $a_{\bf 0}$ ($a^{\dagger}_{\bf 0}$) if $p$ is even or $a_{{\bf k}_0}$ ($a^{\dagger}_{{\bf k}_0}$) if $p$ is odd. This imposes conditions on the parameters of shift \eqref{shift}, which must satisfy now
\begin{eqnarray}
\label{rhog2}
\rho_{-} &=& \rho^*_{-} = \rho^*_{+} = \rho_{+} = \rho e^{-i\phi},\\
\label{rho1_2}
\rho &=& h \frac{1}{2 \sqrt{2}S}\frac{1}{\tilde J_{\bf k_0} - \tilde J_{\bf 0}}
=
h \frac{1}{2 \sqrt{2}S}\frac{1}{J_{\bf 0} - \tilde J_{\bf 0}},
\end{eqnarray}
that implies $\phi = 0$. Notice that Eq.~\eqref{hz2} disappears in this case. Since Eq.~\eqref{h3} does not produce terms proportional to $a_{\bf 0}$ and $a^{\dagger}_{\bf 0}$, no shift of the form \eqref{shift2} is needed. Then, the ground state energy is given by 
\begin{equation}
	\label{gsec2}
	\frac{{\cal E}_h^{(c)}}{N} = S^2 \tilde J_{\bf 0} - \frac{h^2}{4 \left(\tilde J_{{\bf k}_0} - \tilde J_{\bf 0}\right)}
	 = S^2 \tilde J_{\bf 0} - \frac{h^2}{4 \left( J_{\bf 0} - \tilde J_{\bf 0}\right)}
\end{equation}
and there are no higher order field corrections to it. Notice that the negative field correction in Eq.~\eqref{gsec2} is two times smaller than that in the energy \eqref{gse} of the incommensurate state. This circumstance results in a competition between the incommensurate and commensurate phases: although the first term in Eq.~\eqref{gsec2} is greater than that in Eq.~\eqref{gse}, the second term in Eq.~\eqref{gsec2} can compensate this difference and make the commensurate state more energetically favorable at $h>h_c$, where $h_c$ is the critical field. If ${\bf k}_0$ is close to ${\bf g}/2$ in the incommensurate phase, $h_c$ will be small so that our theory based on the expansion in powers of $h$ would remain valid up to $h_c$. We discuss the corresponding phase transition below.

The magnetic order in this phase reads in the second order in $h$ as
\begin{eqnarray}
  \label{oder2k}
	S^{x}_{{\bf r}_j} &=& 
	S \sqrt2\rho,\\
	S^{z}_{{\bf r}_j} &=& 
	S \left(1-\rho^2\right)
	\cos\left({\bf k}_0{{\bf r}_j}\right),\nonumber
\end{eqnarray}
where $\rho$ is given by Eq.~\eqref{rho1_2}.


\subsection{${\bf k}_0={\bf g}/3$}

We turn to a commensurate state in which the vector of the magnetic ordering ${\bf k}_0={\bf g}/3$. Eqs.~\eqref{h3} and \eqref{hz2} produce the following linear terms in the Hamiltonian after performing shift \eqref{shift}:
\begin{eqnarray}
  \label{hz11}
	\mathcal{H}_3^{(1)} = && -i\rho^2S3\sqrt{2NS}\left( J_{\bf 0}-\tilde J_{\bf k_0}\right)
\left( a_{-{\bf k}_0}e^{-2i\phi} - a_{{\bf k}_0}e^{2i\phi} - a^\dagger_{-{\bf k}_0}e^{2i\phi} + a^\dagger_{{\bf k}_0}e^{-2i\phi} \right),\\
\mathcal{H}_{Z2}^{(1)} = && ih\rho\sqrt{NS}\frac12
\left( a_{-{\bf k}_0}e^{-2i\phi} - a_{{\bf k}_0}e^{2i\phi} - a^\dagger_{-{\bf k}_0}e^{2i\phi} + a^\dagger_{{\bf k}_0}e^{-2i\phi} \right),\nonumber
\end{eqnarray}
respectively. Eqs.~\eqref{hz11} give a correction to $\rho_\pm$ in Eq.~\eqref{shift} proportional to $h^2$ which acquire the form (cf.\ Eqs.~\eqref{rhog} and \eqref{rho1})
\begin{eqnarray}
\label{rhocom}
\rho_{-} &=& \rho^*_{+} = \rho e^{-i\phi} - i \rho^2 \frac{3 J_{\bf 0} + 2 \tilde J_{\bf 0} - 5 \tilde J_{{\bf k}_0}}{\sqrt{2}\left(\tilde J_{{\bf k}_0} - \tilde J_{\bf 0}\right)}e^{2 i \phi},
\end{eqnarray}
where $\rho$ is given by Eq.~\eqref{rho1}. This leads also to the correction to the ground state energy proportional to $h^3$ that reads now as (cf.\ Eq.~\eqref{gse})
\begin{equation}
	\label{gsec}
	\frac{{\cal E}_h^{(c)}}{N} = 
	S^2\tilde J_{\bf 0} 
 -\frac{hS\rho}{\sqrt2}
	-hS\rho^2 \frac{3 J_{\bf 0} + 2 \tilde J_{\bf 0} - 5 \tilde J_{{\bf k}_0}}{\tilde J_{{\bf k}_0} - \tilde J_{\bf 0}} \sin3\phi
	+ {\cal O}(h^4).
\end{equation}
Evidently, the energy minimization fixes $\phi$ now. In particular, when DMI is small, it is easy to show that the fraction in the third term in Eq.~\eqref{gsec} is approximately equal to unity and $\sin3\phi$ must be equal to 1. One concludes from Eq.~\eqref{oderh} that third of all spins are antiparallel to the field at such $\phi$. In particular, it is the third term in Eq.~\eqref{gsec} that stabilizes the $Y$ in-plane phase in triangular-lattice Heisenberg antiferromagnets. \cite{chub} It is seen from Eq.~\eqref{gsei4} that the third term in Eq.~\eqref{gsec} can make more preferable the commensurate phase at large enough $h$.

The magnetic ordering reads in the second order in $h$ as
\begin{eqnarray}
  \label{oderhc}
	S^{x}_{{\bf r}_j} &=& 
	S \left[ 
	2\sqrt2\rho (1+2\sqrt2\rho) \cos^2({\bf k}_0{\bf r}_j+\phi) 
	+ 
	\sin({\bf k}_0{{\bf r}_j}+\phi) \left(1-4\rho^2\cos^2({\bf k}_0{\bf r}_j+\phi) \right)
	\right],\\
	S^{z}_{{\bf r}_j} &=& 
	S \left[ 
	-2\sqrt2\rho (1+2\sqrt2\rho) \sin({\bf k}_0{\bf r}_j+\phi)\cos({\bf k}_0{\bf r}_j+\phi) 
	+ 
	\cos({\bf k}_0{{\bf r}_j}+\phi) \left(1-4\rho^2\cos^2({\bf k}_0{\bf r}_j+\phi) \right)
	\right],\nonumber
\end{eqnarray}
where $\phi=\pi/6$ and $\rho$ is given by Eq.~\eqref{rho1}. It can be checked straightforwardly that $(S^{x}_{{\bf r}_j})^2 + (S^{z}_{{\bf r}_j})^2 =S^2(1+{\cal O}(h^3))$.

\subsection{${\bf k}_0={\bf g}/4$}
\label{secg4}

In the commensurate phase in which ${\bf k}_0 = {\bf g}/4$ and $2{\bf k}_0$ is not equal to a reciprocal lattice vector, one has to take into account terms of the order $h^3$ and $h$ in coefficients $\rho_{\pm}$ in Eqs.~\eqref{shift} in order to cancel linear terms in the Hamiltonian proportional to $h^3$ and $h^4$. We suppose in the subsequent consideration that ${\bf k}_0$ in the incommensurate phase is close to ${\bf g}/4$ that can happen only due to frustration of the exchange interaction. Then, DMI, which is normally much smaller than the exchange coupling, plays a minor role in such systems and we neglect it for simplicity assuming that $J_{{\bf k}_0} = \tilde J_{\bf 0}$. In this case, $\gamma_\pm=0$ in Eqs.~\eqref{shift2} and coefficients in shift \eqref{shift} have the form
\begin{eqnarray}
\label{rhocom4k0neq}
\rho_{-} &=& \rho^*_{+} = \rho e^{-i\phi} + (\rho_2 - i\rho_3) e^{i\phi},\\
\label{rhocom4k0eq}
\rho_2 &=& \rho^3 \cos2\phi,\\
\rho_3  &=& \frac12\rho^3 \sin2\phi,
\end{eqnarray}
where $\rho$ is given by Eq.~\eqref{rho1}. The ground state energy has the form
\begin{equation}
    \label{gsec4k0eq}
    \frac{{\cal E}_h^{(c)}}{N} = 
	S^2\tilde J_{\bf 0} 
	- \frac{h^2}{8 \left( \tilde J_{\bf k_0} - \tilde J_{\bf 0} \right)}
        - \frac{h^4 }{512 S^2 \left(\tilde J_{{\bf k}_0} - \tilde J_{\bf 0}\right)^3}\cos4\phi.
\end{equation}
Assuming that ${\bf k}_0 = {\bf g}/4$ differs slightly from $\bf{k}_0$ in the incommensurate state (at which $\tilde J_{\bf 0}$ is minimized), one concludes that the denominator in the third term in Eq.~\eqref{gsec4k0eq} is positive so that the energy is minimized at $\phi=0$.

In this case, the magnetic ordering in the second order in $h$ is given by
\begin{eqnarray}
  \label{oderhc4}
	S^{x}_{{\bf r}_j} &=& 
	S \left[ 
	2\sqrt2 \rho \cos^2({\bf k}_0{\bf r}_j) 
	+ 
	\sin({\bf k}_0{{\bf r}_j}) \left(1-4\rho^2\cos^2({\bf k}_0{\bf r}_j) \right)
	\right],\\
	S^{z}_{{\bf r}_j} &=& 
	S \left[ 
	-2\sqrt2 \rho \sin({\bf k}_0{\bf r}_j)
	\cos({\bf k}_0{\bf r}_j)
	+
	\cos({\bf k}_0{{\bf r}_j}) \left(1-4\rho^2\cos^2({\bf k}_0{\bf r}_j) \right)
	\right].\nonumber
\end{eqnarray}

\section{Transitions between incommensurate and commensurate phases}
\label{transit}

By comparing ground state energies derived above, we calculate in this section critical field values $h_c$ for transitions between the incommensurate phase with ${\bf k}_{0}={\bf k}_{0i}$ to commensurate ones with ${\bf k}_{0}={\bf k}_{0c}$. We will assume that ${\bf k}_{0i}$ is close to ${\bf k}_{0c}$.

\subsection{${\bf k}_{0c}={\bf g}/2$} \label{transit2}

The critical field $h_c$ that stabilizes a commensurate ordering characterized by momentum ${\bf k}_{0c}={\bf g}/2$ can be found by comparing Eqs.~\eqref{gsec2} and \eqref{gse} with the result
\begin{equation}
\label{hc2}
	h_c = 2S \sqrt{2\left(\tilde J^{(c)}_{\bf 0} - \tilde J^{(i)}_{\bf 0} \right) \left(J_{\bf 0} - \tilde J^{(c)}_{\bf 0} \right)},
\end{equation}
where $\tilde J^{(c)}_{\bf 0}$ and $\tilde J^{(i)}_{\bf 0}$ are given by Eq.~\eqref{js} with ${\bf k}_{0}={\bf k}_{0c}$ and ${\bf k}_{0}={\bf k}_{0i}$, respectively.

\subsection{${\bf k}_{0c}={\bf g}/3$}
\label{transit3}

One concludes by comparing Eqs.~\eqref{gsec} and \eqref{gsei4} that due to the third term in Eq.~\eqref{gsec} sufficiently strong field can stabilize a commensurate ordering characterized by momentum ${\bf k}_{0c}={\bf g}/3$ even if $J_{\bf k}$ is minimized on an incommensurate vector ${\bf k}_{0i}$. The critical field $h_c$ of the transition between these phases reads as
\begin{equation}
\label{hc}
	h_c = S32^{1/3} \left(\tilde J^{(c)}_{\bf 0} - \tilde J^{(i)}_{\bf 0} \right)^{1/3} 
	\left(\tilde J_{{\bf k}_{0c}}^{(c)} - \tilde J^{(c)}_{\bf 0} \right)^{2/3},
\end{equation}
where we take into account that the second terms in Eqs.~\eqref{gsec} and \eqref{gsei4} are approximately equal to each other. 
We demonstrate in the next section that it is the transition which arises in $\rm RbFe(MoO_4)_2$.

\subsection{${\bf k}_{0c}={\bf g}/4$}
\label{transit4}

One finds the critical field by comparing Eq.~\eqref{gsei4} with Eq.~\eqref{gsec4k0eq}, the result being
\begin{eqnarray}
\label{hc4}
h_c &=& 4S \left(\tilde J^{(c)}_{\bf 0} - \tilde J^{(i)}_{\bf 0} \right)^{1/4} 
	\left(\tilde J^{(i)}_{{\bf k}_{0i}} - \tilde J^{(i)}_{\bf 0} \right) 
    \left(
    \frac{\tilde J_{2{\bf k}_{0i}}^{(i)} - \tilde J_{\bf 0}^{(i)}}{2\left(J_{2{\bf k}_{0i}} - \tilde J_{\bf 0}^{(i)} \right)^2 }
    \right)^{1/4},
\end{eqnarray}
where we neglect DMI as it is explained in Sec.~\ref{secg4} and take into account that $\tilde J_{2{\bf k}_{0i}}^{(i)} - \tilde J_{\bf 0}^{(i)}$ is small and positive when ${\bf k}_{0i}$ is close to ${\bf k}_{0c}={\bf g}/4$.

We estimate $h_c$ given by Eqs.~\eqref{hc2}--\eqref{hc4} in Appendix~\ref{est} in a simple model of the form \eqref{ham0} and demonstrate that ${\bf k}_0$ changes discontinuously at the IC transition.

\subsection{Quantum and thermal corrections to $h_c$}

It is demonstrated below by the example of $\rm RbFe(MoO_4)_2$ that critical fields $h_c$ can be very sensitive to a small variation of model parameters as well as to quantum and thermal corrections. This sensitivity originates from the first brackets in Eqs.~\eqref{hc2}--\eqref{hc4} which are particularly small at ${\bf k}_{0i}\approx{\bf k}_{0c}$ and gives zero at ${\bf k}_{0i}={\bf k}_{0c}$ (all other terms in these expressions remain finite at ${\bf k}_{0i}={\bf k}_{0c}$). Then, the main contribution to quantum and thermal renormalization of $h_c$ comes from renormalization of the first brackets in Eqs.~\eqref{hc2}--\eqref{hc4} which stem from the difference of the ground state energies of commensurate and incommensurate phases at $h=0$. Then, one has to consider quantum and thermal corrections to ${\cal E}_0$ at $h=0$ which can be easily found from Eqs.~\eqref{h2} and \eqref{spec} and which have the form in the first order in $1/S$
\begin{eqnarray}
\label{de}
\delta{\cal E}_0 &=& 
	\frac{1}{2}\sum_{\bf k} 
	\left( \epsilon_{\bf k}(1+2n_{\bf k} )-
	S \left( J_{\bf k} + \tilde J_{\bf k} - 2 \tilde J_{\bf 0} + \tilde A \right) \right).
\end{eqnarray}
As a result, one has to make the following replacement in first brackets of Eqs.~\eqref{hc2}--\eqref{hc4}:
\begin{equation}
\label{dhc}
\tilde J^{(c)}_{\bf 0} - \tilde J^{(i)}_{\bf 0}
\mapsto
\tilde J^{(c)}_{\bf 0} - \tilde J^{(i)}_{\bf 0}
+\frac{1}{S^2N} \left( \delta{\cal E}^{(c)}_0 - \delta{\cal E}^{(i)}_0 \right),
\end{equation}
where $\delta{\cal E}^{(c)}_0$ and $\delta{\cal E}^{(i)}_0$ are given by Eq.~\eqref{de} at ${\bf k}_0={\bf k}_{0c}$ and ${\bf k}_0={\bf k}_{0i}$, respectively.

\section{Application to $\rm RbFe(MoO_4)_2$}
\label{app}

\subsection{General consideration of the model}

\begin{figure}
\includegraphics[scale=0.5]{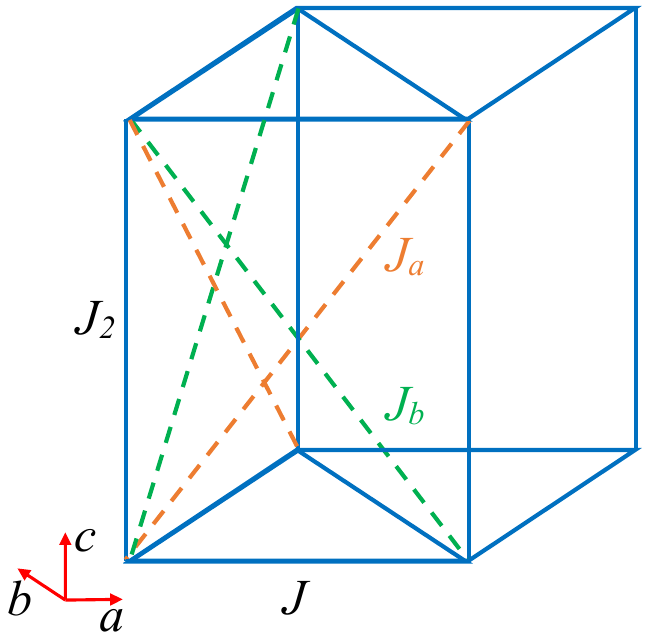}
\caption{
Schematic view of the magnetic lattice formed by Fe$^{3+}$ ions in $\rm RbFe(MoO_4)_2$ in which triangular-lattice $ab$ planes are stacked along $c$ axis. Exchange coupling constants of the model are indicated. Axis $c$ is the three-fold symmetry axis.
\label{lattice}}
\end{figure}

\begin{figure}
\includegraphics[scale=0.3]{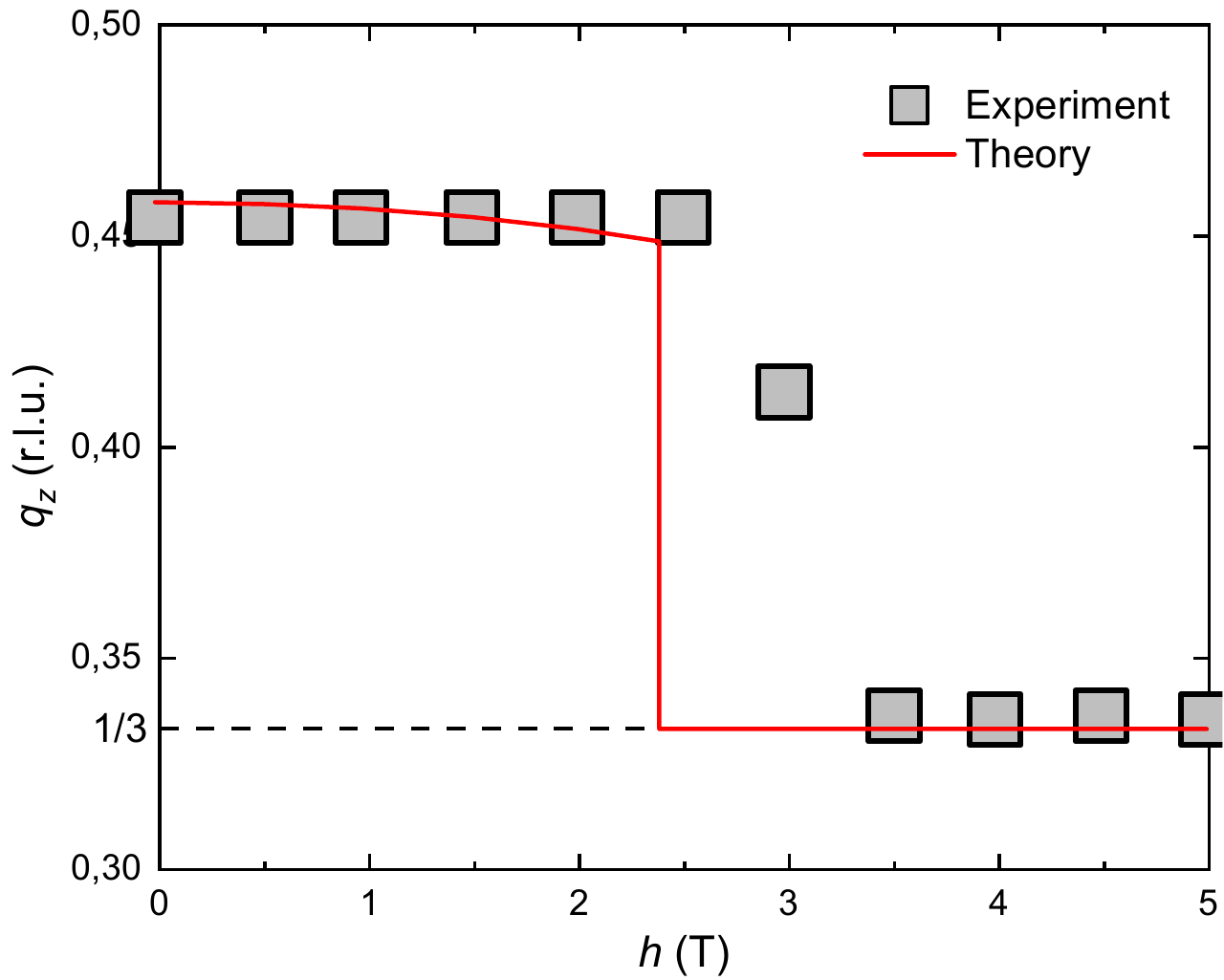}
\caption{
Component $q_z$ of the vector of the magnetic structure ${\bf k}_0=(1/3,1/3,q_z)$ in $\rm RbFe(MoO_4)_2$. Neutron data are taken from Fig.~2(d) of Ref.~\cite{rbfemoo4}. Theoretical curve at $h<h_c\approx2.4$~T is given by Eq.~\eqref{qzh}.
\label{qzfig}}
\end{figure}

It was well established before \cite{rbfemoo1,rbfemoo2,rbfemoo3,rbfemoo4,rbfemoo5} that $\rm RbFe(MoO_4)_2$ is a layered spin-$\frac52$ triangular-lattice antiferromagnet described by model \eqref{ham0} whose magnetic lattice is depicted in Fig.~\ref{lattice}. At small field, an incommensurate magnetic ordering with ${\bf k}_0 = {\bf k}_{0i} = (1/3,1/3,q_z)$ is stabilized by weak competing inter-plane exchange interactions. \cite{rbfemoofp} The magnetic structure switches from ${\bf k}_{0i}$ to ${\bf k}_0 = {\bf k}_{0c} = (1/3,1/3,1/3)$ in a small interval of the in-plane field and it remains commensurate upon further field increasing (see Fig.~\ref{qzfig}). However the position of the center of this interval varies significantly from $3.0$~T to $3.8$~T across different experiments. \cite{rbfemoo1,rbfemoo3,rbfemoo4} Values of 0.44, 0.45, and 0.458 were reported for $q_z$ in Ref.~\cite{rbfemoomult1}, Refs.~\cite{mitamura,inami}, and Ref.~\cite{rbfemoo4}, respectively. Then, the results of the above theoretical consideration can be applied to $\rm RbFe(MoO_4)_2$.

To the best of our knowledge, a comprehensive analysis has not been carried out yet for determination of all model parameters describing $\rm RbFe(MoO_4)_2$. For instance, magnon spectra found by inelastic neutron scattering in Ref.~\cite{rbfemoo5} were discussed using a simplified model in which inter-plane interaction is described by only one effective exchange coupling constant $J_2$ shown in Fig.~\ref{lattice} (which by itself cannot lead to the spiral). To reproduce the incommensurate ordering along $c$ axis, we adopt the model of the inter-plane interaction suggested in Refs.~\cite{rbfemoofp,rbfemoomult1} which involves three exchange coupling constants $J_2$, $J_a$, and $J_b$ depicted in Fig.~\ref{lattice}. Besides, we find below that taking into account dipolar forces which are unavoidable in real materials improves the agreement of $h_c$ with experimental findings while their influence on magnon spectra is quite small in $\rm RbFe(MoO_4)_2$. The Hamiltonian of dipolar interaction has the form
\begin{equation}
\label{hamdip}
\mathcal{H}_d = -\frac{1}{2}\sum_{l\neq m} Q_{lm}^{\alpha \beta} S_l^\alpha S_{m}^\beta,
\end{equation}
where
\begin{equation}
\label{dip_forces}
Q_{lm}^{\alpha \beta} = \frac{\omega_0}{4\pi}\frac{3R_{lm}^\alpha R_{lm}^\beta-\delta_{\alpha \beta}R_{lm}^2}{R_{lm}^5} 
\end{equation}
is the dipolar tensor,
\begin{equation}
	\omega_0 = 4\pi \frac{(g\mu_B)^2}{v_0} \approx 0.013 \mbox{ meV},
\end{equation}
$v_0$ is the unit cell volume, and $\omega_0$ is the characteristic dipolar energy whose value is given for $\rm RbFe(MoO_4)_2$. It can be shown that the dipolar interaction leads to the following renormalization of parameters in expressions for the ground state energy and the bilinear part of the Hamiltonian \eqref{h2}:
\begin{eqnarray}
\label{renorm}
J_{\bf k} &\mapsto& J_{\bf k} - \frac12 Q^{yy}_{\bf k},\nonumber\\
\tilde J_{\bf k} &\mapsto& \tilde J_{\bf k} + \frac14 Q^{yy}_{{\bf k}+{\bf k}_0}.
\end{eqnarray}
To be precise, umklapp terms also arise in ${\cal H}_2$ having the form $a_{\bf k}a_{{\bf k}\pm{\bf k}_0}$,  $a^\dagger_{\bf k}a_{{\bf k}\pm{\bf k}_0}$, and $a^\dagger_{\bf k}a^\dagger_{{\bf k}\pm{\bf k}_0}$ which are proportional to $Q^{yx}_{\bf k}$ and $Q^{yz}_{\bf k}$ and which greatly complicate the analysis. However umklapp terms have little influence on magnon spectra and lead to a very small splitting of magnon branches \cite{batalov} which are not seen in available neutron data in $\rm RbFe(MoO_4)_2$. Then, we neglect umklapp terms below. Dipolar sums computation technique \cite{cohen} should be used to find $Q^{yy}_{\bf k}$ in Eqs.~\eqref{renorm}.

In particular, we have
\begin{equation}
\label{jk0}
	\tilde J_{\bf 0}^{(i)} = 
	-\frac32J + J_2 \cos(2\pi q_z) 
	-\frac32 (J_a+J_b) \cos(2\pi q_z) 
	-\frac{3\sqrt3}{2}(J_a-J_b)\sin(2\pi q_z)
	 + \frac14 Q^{yy}_{{\bf k}_{0i}}.
\end{equation}
Minimization of Eq.~\eqref{jk0} gives for $q_z$ \cite{rbfemoofp,rbfemoomult1}
\begin{equation}
\label{tan}
	\tan(2\pi q_z) = \frac{3\sqrt3(J_a-J_b)}{3(J_a+J_b) -2J_2},
\end{equation}
where we take into account that according to our calculations  $Q^{yy}_{{\bf k}_{0i}}$ weakly depends on $q_z$ and one can neglect the very small dipolar contribution in Eq.~\eqref{tan}.

Using this model, we revisit first previous experimental data to find the optimal set of parameters which successfully describes the majority of them.

\subsection{Choice of model parameters}

As it is demonstrated below, this set of parameters expressed in meV has the form
\begin{eqnarray}
\label{param}
	J &=& 0.0872, \nonumber\\
	A &=& 0.0319, \nonumber\\
	J_2 &=& 0.0185, \\
	J_a &=& 0.0057, \nonumber\\
	J_b &=& 0.0055. \nonumber
\end{eqnarray} 
Notice that $J$ in Eq.~\eqref{param} coincides with previously established value of $0.086(2)\,{\rm meV}$. Quantum renormalization of the value of $A$ from Eq.~\eqref{param} described by Eq.~\eqref{at} gives 0.026~meV in agreement with previously found result $0.027(1)$~meV for the magnitude of the easy-plane anisotropy. However, $J_2$ in Eq.~\eqref{param} is approximately an order of magnitude larger than the single effective inter-plane coupling $J'$ suggested before (which corresponds to our $J_2$ and which by itself cannot lead to the spiral along $c$ axis). \cite{rbfemoo1,rbfemoo2,rbfemoo3,rbfemoo4,rbfemoo5} The smallness of $J'$ reflects the frustrating nature of the inter-plane interaction in $\rm RbFe(MoO_4)_2$ and that is why our $J_2$ is an order of magnitude larger than $J'$. However, $J_2$ value in Eq.~\eqref{param} is in a quantitative agreement with results of the first-principles calculations \cite{rbfemoofp} and $J_{a,b}$ values are consistent with the first-principles results in order of magnitude.

It can be checked using Eq.~\eqref{tan} that parameters \eqref{param} give $q_z=0.458$ in agreement with experimental findings \cite{mitamura,inami,rbfemoo4} and $J_{{\bf k}_{0c}}<J_{{\bf k}'_{0c}}$, where ${\bf k}'_{0c}=(1/3,1/3,-1/3)$. 

Values of the saturation field in model \eqref{ham0} for field directions parallel and transverse to the anisotropy axis have the form
\begin{eqnarray}
\label{hsp}
h_s^\| &=& 2S\left(J_{\bf 0} - J_{{\bf k}_{0i}} + \tilde A -\frac12 Q^{cc}_{\bf 0} -\frac14 Q^{cc}_{{\bf k}_{0i}}\right),\\
\label{hst}
h_s^\perp &=& 2S\left(J_{\bf 0} - J_{{\bf k}_{0i}} -\frac12 Q^{aa}_{\bf 0} -\frac14 Q^{aa}_{{\bf k}_{0i}}\right),
\end{eqnarray}
respectively, where we take into account that $Q^{aa}_{\bf 0}=Q^{bb}_{\bf 0}$ and $Q^{aa}_{{\bf k}_{0i}}=Q^{bb}_{{\bf k}_{0i}}$. Eqs.~\eqref{param}, \eqref{hsp}, and \eqref{hst} give $h_s^\|=20.3$~T and $h_s^\perp=19.2$~T which agree with respective experimental results \cite{rbfemoo2} 19.9~T and 18.2(2)~T (dipolar contributions to the theoretical values are $-0.11$~T and $-0.08$~T, respectively). Notice that $Q^{tt}_{\bf 0} = \omega_0(1/3-{\cal N}_t)$, where ${\cal N}_t$ is the demagnetization tensor component in the direction $t$ which we set equal to zero in all calculations (thus assuming that the sample is infinite in the corresponding direction).

The frequency $\Omega$ of antiferromagnetic resonance at zero field is given in the linear spin-wave approximation by Eq.~\eqref{spec} at ${\bf k = k}_{0i}$. Taking into account quantum and thermal corrections to this finding, one has to replace $S$ by its renormalized value $\overline{S}$, \cite{chub} the result being
\begin{eqnarray}
	\label{afr}
	\Omega &=& 2\overline{S}\sqrt{\left( \tilde J_{{\bf k}_{0i}} - \tilde J_{\bf 0} \right) \tilde A},\\
	\label{st}
	\overline{S} &=& S - \frac1N\sum_{\bf k} \frac{S \left( J_{\bf k} + \tilde J_{\bf k} - 2 J_{{\bf k}_{0i}} + \tilde A \right)(1+2n_{\bf k}) - \epsilon_{\bf k}}{2\epsilon_{\bf k}},
\end{eqnarray}
where $n_{\bf k} = (e^{\epsilon_{\bf k}/T}-1)^{-1}$ is the Plank's function and renormalization \eqref{renorm} is also implied. One obtains from these equations $\overline{S} = S-0.136$ and $\Omega=87$~GHz at $T=0$. The latter value is close to 90~GHz obtained by ESR measurements \cite{rbfemoo2}. Values of the staggered magnetization found in neutron scattering experiments 1.95(25) \cite{mitamura} and 2.10(5) \cite{inami} are somewhat smaller than the value of 2.36 predicted by Eq.~\eqref{st} at $T=0$. However, taking into account thermal fluctuations at the temperature of experiments brings the theoretical result $\overline{S}=2.26$ closer to the experimental findings.

\begin{figure}
\includegraphics[scale=0.4]{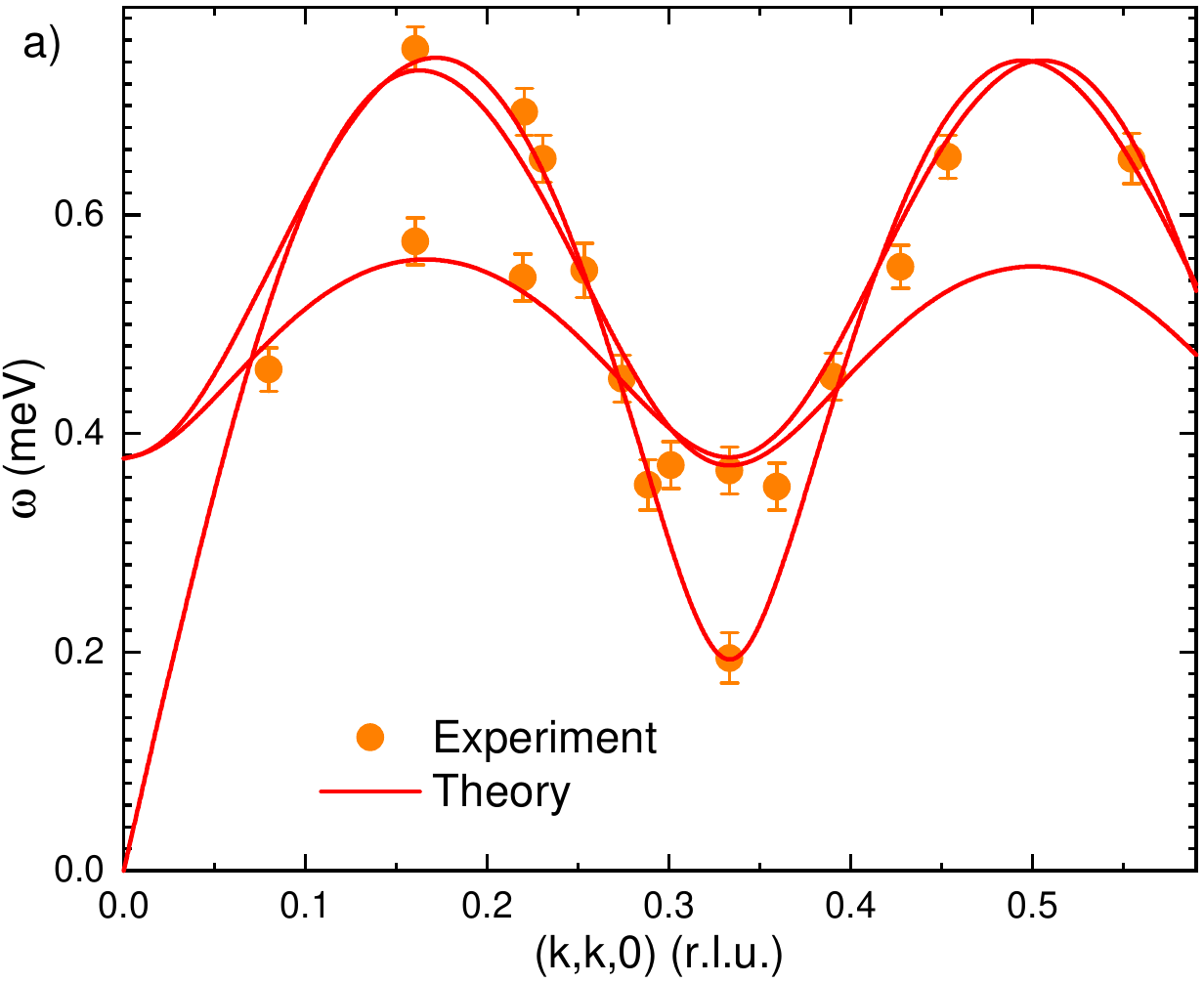}
\includegraphics[scale=0.4]{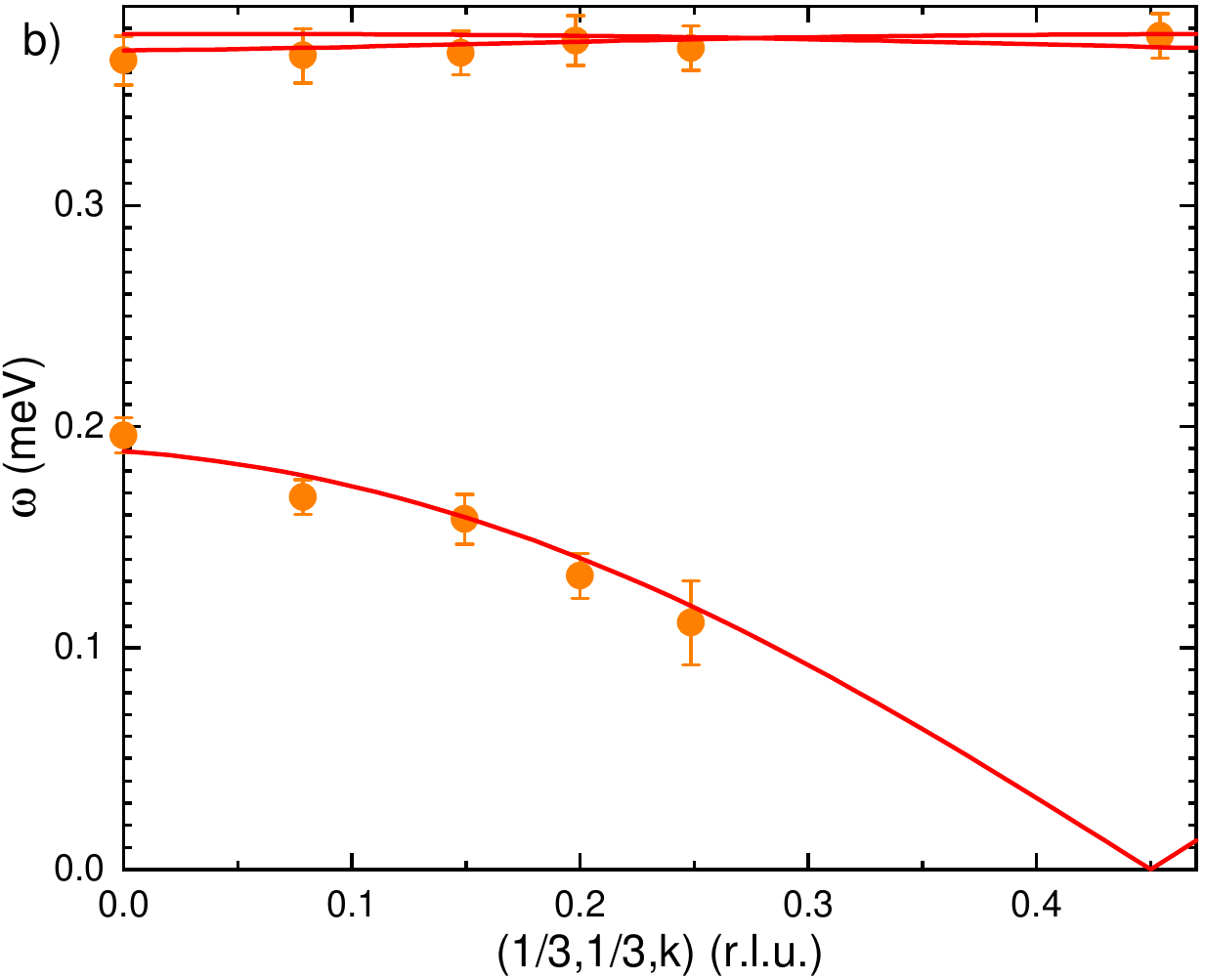}
\caption{
Magnon spectra extracted from the neutron scattering experiment in $\rm RbFe(MoO_4)_2$ (data are taken from Figs.~2(c) and 2(d) of Ref.~\cite{rbfemoo5}) and calculated in the linear spin-wave theory using Eq.~\eqref{spec}. Three branches in the theory correspond to $\epsilon_{\bf k}$, $\epsilon_{\bf k+k_0}$, and $\epsilon_{\bf k-k_0}$.
\label{neutronfig}}
\end{figure}

It is seen from Fig.~\ref{neutronfig} that spectra of three magnon branches corresponding to $\epsilon_{\bf k}$, $\epsilon_{\bf k+k_0}$, and $\epsilon_{\bf k-k_0}$ and calculated using Eqs.~\eqref{spec}, \eqref{renorm}, and \eqref{param} describe well neutron scattering data from Ref.~\cite{rbfemoo5}. The simplified model with $J_a=J_b=0$ proposed in Ref.~\cite{rbfemoo5} (which does not, however, explain the helical ordering along $c$ axis) also reproduces well neutron spectra using Eq.~\eqref{spec} at ${\bf k}_0={\bf k}_{0i}$ and $J_2=0.0017$~meV (but not at $J_2=0.0007$~meV as reported in Ref.~\cite{rbfemoo5}).

Parameters \eqref{param} are obtained by setting the component $q_z$ of the ordering vector to 0.458 (Eq.~\eqref{tan}) and by simultaneous finding of reasonably good fits of experimentally obtained magnon spectra (Fig.~\ref{neutronfig}), saturation fields given by Eqs.~\eqref{hsp} and \eqref{hst}, and $h_c$ value (Eqs.~\eqref{hc} and \eqref{dhc}).

\subsection{Low-field behavior of $\rm RbFe(MoO_4)_2$}

Eqs.~\eqref{hc} and \eqref{param} give $h_c\approx2.1$~T which is quite far from the experimental value of $\approx3$~T \cite{rbfemoo1,rbfemoo4,rbfemoo3}. We point out, however, that $h_c$ given by Eq.~\eqref{hc} is extremely sensitive to small changes of model parameters in $\rm RbFe(MoO_4)_2$. In particular, taking into account dipolar forces changes $h_c$ from 1.9~T to 2.1~T (unlike other physical observables considered above which depend weakly on the dipolar interaction). 

This sensitivity originates from the first bracket in Eq.~\eqref{hc} which is particularly small and gives zero at ${\bf k}_{0i}={\bf k}_{0c}$ (the second bracket in Eq.~\eqref{hc} remains finite in this case). As it is explained above, renormalization of the first bracket in Eq.~\eqref{hc} in the first order in $1/S$ is given by Eq.~\eqref{dhc}. Taking into account these corrections gives $h_c\approx2.25$~T at $T=0$, and $h_c\approx2.4$~T at $T=2.8$~K (the temperature of the experiment in Ref.~\cite{rbfemoo4} which results are shown in Fig.~\ref{qzfig}).

\begin{figure}
\includegraphics[scale=0.34]{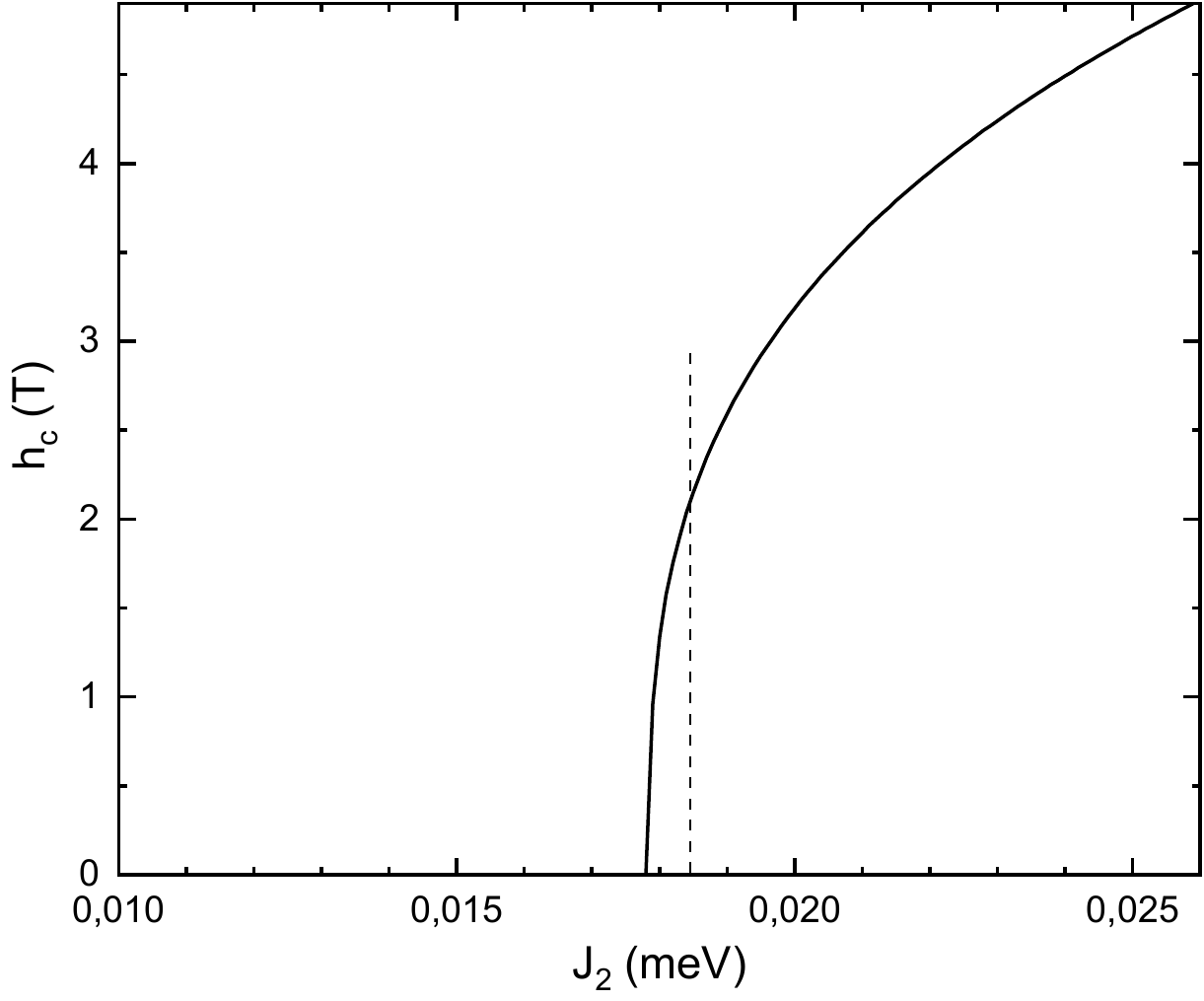}
\caption{
Dependence of the critical field $h_c$ given by Eq.~\eqref{hc} on $J_2$ when all other model parameters are fixed to their values from Eq.~\eqref{param}. The vertical dashed line marks $J_2$ value from Eq.~\eqref{param}.
\label{hcfig}}
\end{figure}

Notice also that quantum and thermal corrections to the magnon spectrum which was measured at a finite temperature would also change slightly parameters \eqref{param} after the fitting procedure. 
Although we expect that taking into account the magnon-magnon interaction 
\footnote{as well as some other possible minor spin-spin and spin-lattice couplings} 
would not lead to a large change of exchange coupling constants due to the large spin value $S=5/2$, it could change noticeably $h_c$ due to the great sensitivity of this quantity to small variation of the model parameters in $\rm RbFe(MoO_4)_2$.
Fig.~\ref{hcfig} illustrates this sensitivity and demonstrates that an increasing of $J_2$ by a few percents would increase $h_c$ by several tenths Tesla. Calculation of the magnon spectrum in the first order in $1/S$ is a cumbersome task in noncollinear magnets which is out of the scope of the present paper. We find reasonable the agreement of obtained values of $h_c$ with previous experimental data and restrict ourselves to the consideration carried out.

We point out that $\tilde J_{\bf k}$ is not quadratic in $\rm RbFe(MoO_4)_2$ near ${\bf k}={\bf k}_{0i}$. Then, $q_z$ should depend on $h$ as it is explained in Sec.~\ref{sf}. Particular calculations show that
\begin{equation}
\label{qzh}
	q_z = 0.458 - 0.59 \left( \frac{h}{h_s^\perp}\right)^2.
\end{equation}
However this dependence is very weak as it is shown in Fig.~\ref{qzfig}.

Substituting parameters \eqref{param} to Eqs.~\eqref{rho1} and \eqref{gamma} one obtains
\begin{eqnarray}
\label{rgexp}
	\rho &\approx& 0.0364h,\nonumber\\
	\gamma &\approx& 0.684\rho^2,
\end{eqnarray}
where it is assumed that $h$ is expressed in Tesla. Remember that the expansion in powers of $h$ carried out in Sec.~\ref{sf} is justified if $\gamma\ll\rho$ which can be written using Eq.~\eqref{rgexp} as $0.7\rho\ll1$. The latter inequality holds well for $h<4$~T.

\section{Summary and conclusion}
\label{conc}

We discuss field-induced transitions between phases with incommensurate and commensurate magnetic orderings (IC transitions) in easy-plane helical antiferromagnets described by model \eqref{ham0}. We consider the magnetic field $\bf h$ applied in the easy plane which is much smaller than the saturation field $h_s$. This smallness allows us to find analytical expressions for physical quantities as series in $h$. The incommensurate spiral is assumed to appear in the zero field due to small DMI and/or frustration of the exchange spin couplings. We demonstrate that the field distorts the helix and the resultant spin ordering is described by a set of harmonics of the vector of the magnetic structure ${\bf k}_0$. This ordering is given by Eq.~\eqref{oderhi} in the second order in $h$ which can be written in the more compact form \eqref{order22}. It is demonstrated that the field can gradually change the vector of the magnetic structure ${\bf k}_0$  as it was observed in some experiments (see, e.g., Refs.~\cite{rbfemoo4,mitamura}).

We assume that ${\bf k}_0$ is close to ${\bf g}/n$, where ${\bf g}$ is a reciprocal lattice vector and $n$ is integer. We show that upon field increasing the ground state energy of the commensurate phase whose magnetic structure is described by the vector ${\bf g}/n$ becomes lower than the energy of the incommensurate phase with the distorted spiral. We consider particular cases of $n=2$, 3, and 4 and find Eqs.~\eqref{hc2}, \eqref{hc}, and \eqref{hc4} for critical fields $h_c$ of corresponding first-order IC phase transitions. Then, the magnetic ordering changes abruptly at $h=h_c$ from that described by Eq.~\eqref{oderhi} to those given by Eqs.~\eqref{oder2k}, \eqref{oderhc}, and \eqref{oderhc4}. We take into account also quantum and thermal corrections to critical fields (see Eq.~\eqref{dhc}). The observed first-order character of IC transitions is in agreement with consideration in Ref.~\cite{martynov} of the case of $n=2$. We do not consider the cases of $n>4$ because the complexity of calculations increases as $n$ rises. Besides, as far as we know, transitions to commensurate phases with $n\ge4$ haven't been detected experimentally so far. 


Application of our theory at $n=2$ to particular material NdFe$_{3}$(BO$_{3}$)$_{4}$ is discussed in Ref.~\cite{boratei} in relation with corresponding neutron experiments. In the present paper, we apply our results for $n=3$ to triangular-lattice material $\rm RbFe(MoO_4)_2$. \cite{rbfemoofp} As a by-product of the main consideration, we find model parameters which describe more accurately the full set of available experimental data suggested before for $\rm RbFe(MoO_4)_2$.

We point out that, unlike most of preceding works regarding IC transitions in helical magnets,\cite{dzya64, Izyumov1984, Nagamiya1962, maslov} our analysis of the classical ground state does not employ the continuous approximation. Our results should be adequate if $h_c\ll h_s$ and their generality provides possibilities of further experimental verification.

We carry out also a general consideration of the IC transition in the continuous approximation in a simple classical model similar to that discussed by Dzyaloshinskii in Ref.~\cite{dzya64} and show that it is governed by a sine-Gordon equation for arbitrary $n\ge2$ (see Appendix~\ref{clim_arb} and Eq.~\eqref{sG_arb}).

\acknowledgments

We are grateful to L.E.\ Svistov for fruitful discussions.

\appendix

\section{Simple model}
\label{est}

We provide in this appendix quantitative estimations for some claims made in the main text. We employ for this purpose a simple model \eqref{ham0} of a classical spin chain with coupling constants $J_1$ and $J_2$ between nearest and next-nearest spins, respectively. We also assume for simplicity that there is no DMI and that the easy-plane anisotropy is strong enough to confine spins to the $xz$ plane. In this case, Hamiltonian \eqref{ham0} has the form
\begin{equation}
 \mathcal{H} = 
\sum_{j} 
\Bigl(
J_{1} \left({\bf S}_j {\bf S}_{j+1}\right) + J_{2} \left({\bf S}_j {\bf S}_{j+2}\right) \Bigr)
+
A\sum_{j} \left(S^y_j\right)^2,
\end{equation}
where all lattice vectors are one-dimensional. Then, the magnetic structure vector at zero field given by minimization of the ground state energy $\frac{{\cal E}_0}{N S^2} = \tilde{J}_0 = J_1 \cos k_0 + J_2 \cos 2k_0$ is $k_{0 i}^{(0)} = \arccos \left( -\frac{J_1}{4 J_2} \right)$. 

Let us consider the case of the transition to the commensurate phase with ${\bf k}_{0 c} = {\bf g}/2$ (Sec.~\ref{transit2}). In order ${\bf k}_{0 i}^{(0)}$ to be close to ${\bf k}_{0 c}$, we choose parameters $J_1$ and $J_2$ so that 
\begin{equation}
\alpha = 1 - \frac{J_1}{4J_2} \ll 1. 
\end{equation}
Then, $k_{0 i}^{(0)} = \pi-\sqrt{2 \alpha} + {\cal O}(\alpha^{3/2})$. The critical field given by Eq.~\eqref{hc2} has the form
\begin{equation}
\label{hc1d}
    h_c = 2 \sqrt{2} \alpha S J_1.
\end{equation}

The vector of magnetic structure depends on the magnetic field as it is discussed in Sec.~\ref{sf}: $k_0 (h) =  k_{0 i}^{(0)} + \delta k_0$.
We obtain the field correction $\delta k_0$ in the leading order in $h$ by minimizing the ground state energy \eqref{gse}, the result being
\begin{equation}
    \delta k_0 (h) = -\frac{h^2}{8 \sqrt{2} S^2 J_1^2 \alpha^{1/2}}
\end{equation}
that gives at $h = h_c$
\begin{equation}
    \delta k_0 (h_c) = -\frac{\alpha^{3/2}}{\sqrt{2}}\ll1.
\end{equation}
Then, the magnetic field induces only a small correction to the magnetic structure vector in the incommensurate phase and does not interfere with the requirement that $k_{0 i} \approx k_{0 c}$. The jump of the magnetic structure vector is finite at the transition:
\begin{equation}
 k_{0 c} - k_{0 i} (h_c) = \sqrt{2 \alpha}+ \frac{7 \alpha^{3/2}}{6 \sqrt{2}} + {\cal O}(\alpha^2).
\end{equation}

Let us turn to the transition to the phase with $k_{0 c} = g/3$ (see Sec.~\ref{transit3}). The parameter which needs to be small in order for $k_{0 i}^{(0)}$ to be close to $k_{0 c}$ is
\begin{equation}
			\alpha = \frac{1}{2} - \frac{J_1}{4 J_2} \ll 1.
\end{equation}
The critical field given by Eq.~\eqref{hc} is
\begin{equation}
\label{hc1d3}
    h_{c 3} = \left(\frac{81}{2}\right)^{1/3} \alpha^{2/3} S J_1
\end{equation}
which is larger than $h_c$ given by Eq.~\eqref{hc1d} at the same $J_1$ and $\alpha\ll1$.

In the case of the transition to a phase with $k_{0 c} = g/4$, the requirement $k_{0 i}^{(0)} \sim k_{0 c}$ implies 
\begin{equation}
	\alpha = \frac{J_1}{4 J_2} \ll 1.
\end{equation}
 Then, Eq. \eqref{hc4} gives
\begin{equation}
    h_{c 4} = 2 \sqrt{2} S J_1
\end{equation}
which is much smaller than the saturation field $h_s \sim J_2$ and which is much larger than critical fields given by Eqs.~\eqref{hc1d} and \eqref{hc1d3} at the same $J_1$ and $\alpha\ll1$.



\section{Continuous limit for $n=3$}
\label{clim}

We consider in this appendix a simple model in which the IC transition arises to the phase with ${\bf k}_{0c}={\bf g}/3$ (i.e., at $n=3$) and show that the system is governed by the sine-Gordon equation in the continuous limit in the leading order in $h$. We study the simplest version of model \eqref{ham0} describing a classical spin chain with coupling constants $J=1$ and the DMI between nearest spins. In this case, Hamiltonian \eqref{ham0} has the form
\begin{equation}
\label{ham02}
 \mathcal{H} = 
\sum_{j} 
\Bigl(
\left({\bf S}_j {\bf S}_{j+1}\right) + {\bf D} \left[{\bf S}_j \times {\bf S}_{j+1}\right] - hS^x_j\Bigr)
+
A\sum_{j} \left(S^y_j\right)^2,
\end{equation}
Our consideration below is an extension of that carried out in model \eqref{ham02} in Ref.~\cite{maslov} at $D\ll1$ (in which case the IC transition occurs to the phase with ${\bf k}_{0c}={\bf g}/2$). 

Introducing angles $\theta_i$ of spins, one has for the classical ground state energy
\begin{equation}
\label{e0a}
	{\cal E}_0 = -\sum_j \left(\sqrt{1+D^2}S^2\cos\left(\theta_j-\theta_{j+1}+\alpha\right) 
	+ hS \cos\theta_j\right),
\end{equation}
where $\alpha = \pi+\arctan D$. At $h=0$, the angle between nearest spins $\theta_{j+1}-\theta_j=\alpha=k_{0 i}^{(0)}$ is obtained by minimization of the ground state energy \eqref{e0a}. Assuming that $\alpha=k_{0 i}^{(0)}\approx2\pi/3$ (i.e., $D\sim1$ and ${\bf k}_{0i}\approx {\bf g}/3$) at finite $h$, it is convenient to make the replacement $\theta_j=\varphi_j+2\pi j/3$ so that one has from Eq.~\eqref{e0a}
\begin{equation}
\label{e0a2}
	{\cal E}_0 = -\sum_j \left(\sqrt{1+D^2}S^2\cos\left(\varphi_j-\varphi_{j+1}+\alpha-\frac{2\pi}{3}\right) 
	+ hS \cos\left(\varphi_j+\frac{2\pi}{3}j\right)\right).
\end{equation}
Using the smallness of the argument of the first cosine in Eq.~\eqref{e0a2}, one obtains after minimization of the ground state energy
\begin{equation}
\label{mina1}
	\varphi_{j-1} + \varphi_{j+1} - 2\varphi_j = \tilde h \sin\left(\varphi_j+\frac{2\pi}{3}j\right), \qquad \forall j,
\end{equation}
where $\tilde h = h/S\sqrt{1+D^2}$. Due to the term $2\pi j/3$ in Eq.~\eqref{mina1}, it is convenient to divide the lattice into three sublattices and to introduce three slowly varying functions $\Phi(j)$, $y_{1,2}(j)$ so that $\varphi_j=\Phi(j)$, $\varphi_{j}=\Phi(j)+y_1(j)$, and $\varphi_{j}=\Phi(j)+y_2(j)$ on the first, on the second, and on the third sublattices, respectively. As a result, one obtains from Eq.~\eqref{mina1} in the continuous limit
\begin{equation}
\label{set}
\left\{
	\begin{aligned} 
		\Phi'' + y_1+y_2+y_1'-y_2'+\frac12(y_1''+y_2'') &=\tilde h \sin\Phi,\\
		\Phi'' - 2y_1+y_2+y_2'+\frac12y_2'' &=\tilde h \sin\left(\frac{2\pi}{3}+\Phi+y_1\right),\\
		\Phi'' + y_1-2y_2-y_1'+\frac12y_1'' &=\tilde h \sin\left(\frac{4\pi}{3}+\Phi+y_2\right).
	\end{aligned}
	\right.
\end{equation}
One has in the leading order in $\tilde h$ from Eqs.~\eqref{set} after adding all equations together and using the fact that $y_{1,2}\sim h$, $y_{1,2}'=O(h^2)$, and $y_{1,2}''=O(h^3)$
\begin{equation}
\label{phi}
	\Phi''+\frac13\left(y_1''+y_2''\right)=\frac{1}{24}\tilde h^3\sin3\Phi.
\end{equation}
Expressing $\Phi''$ from the first equation \eqref{set}, substituting the result to the last two equations in \eqref{set}, and adding the last two equations together one gets $y_1+y_2=\tilde h\sin\Phi$. Substituting this result to Eq.~\eqref{phi} and denoting $\tilde\Phi=\Phi+\frac13\tilde h\sin\Phi$, one obtains the sine-Gordon equation in the leading order in $\tilde h$
\begin{equation}
\label{phi2}
	\tilde\Phi''=\frac{1}{24}\tilde h^3\sin3\tilde\Phi.
\end{equation}
At $D\ll1$, when the IC transition occurs to the phase with ${\bf k}_{0c}={\bf g}/2$, one comes to the sine-Gordon equation with $\tilde h^2\sin2\tilde\Phi$ in the right side of Eq.~\eqref{phi2}. \cite{maslov} It was shown in Ref.~\cite{martynov} that higher order terms in $\tilde h$ (containing higher order derivatives) which are ignored in Eq.~\eqref{phi2} make the IC transition discontinuous.

\section{Continuous limit for arbitrary $n$}
\label{clim_arb}

In this appendix we extend the approach of Appendix \ref{clim} to transitions to the phase with ${\bf k}_{0c}={\bf g}/n$, where $n\geq 2$ is arbitrary. We show below that the system is governed by the sine-Gordon equation in the leading order in $h$, with the rate of change of the spin angles (square root of the coefficient on the right hand side of Eq.~\eqref{phi2} and its generalizations) proportional to $h^{n/2}$.

Just as in Appendix \ref{clim}, we study a classical spin chain described by Hamiltonian \eqref{ham02} with the coupling constant $J=1$ and the DMI with absolute value $D$. We assume that the relation between $D$ and $J$ is such that the angle between spins at zero field $\alpha = \pi + \arctan D$ is close to $2 \pi / n$. If angles of spins are denoted by $\theta_j$ as before, then the replacement $\theta_j = \varphi_j + 2 \pi j/n$ allows to write the energy minimization conditions in the form 
\begin{equation}
\label{min_arb}
	\varphi_{j-1} + \varphi_{j+1} - 2\varphi_j = \tilde h \sin\left(\varphi_j+\frac{2\pi}{n}j\right), \qquad \forall j,
\end{equation}
where $\tilde h = h/S\sqrt{1+D^2}$ and we have taken advantage of the fact that $\varphi_{j+1} - \varphi_j - \alpha+ \frac{2\pi}{n} \ll 1$ in order to linearize the left hand side. Below we sketch the process of constructing the solutions to the system of difference equations \eqref{min_arb}. To lighten notations, we omit the tilde in $\tilde h$ below.

Let us try the solutions in the form $\varphi_k = \Phi(k) + y_j (k)= \Phi(k) + \sum_{m=1}^n y_{j, m}(k)$, where $\Phi, y_{j,m}$ are slowly changing functions, $j = k \mod n$ (so that we only have to consider equations \eqref{min_arb} for $j = 0 \ldots n-1$), and $y_{j, m} \sim h^m$. Then, Eq.~\eqref{min_arb} takes the form in the continuous limit
\begin{equation}
    \label{phiy}
    \Phi'' + y_{j+1} - 2 y_j + y_{j-1} + \sum\limits_{l=1}^\infty \frac{y_{j+1}^{(l)} + (-1)^l y_{j-1}^{(l)}}{l!} = h \sin \left( \Phi + y_j + \frac{2 \pi j}{n} \right), \quad j = 0 \ldots n-1,
\end{equation}
where $y_{j}^{(l)}$ denotes the $l$-th derivative. Considering powers of $h$ in these equations, we construct $y_{j, m}$ as functions of $\Phi$. The first iteration of perturbation theory has a solution $y_{j, 1} = \frac{h \sin(\Phi + \frac{2 \pi j}{n})}{2 (\cos \frac{2 \pi}{n}-1)}$ (here we assume that $\Phi'$ has the order $1$ or higher in $h$ so that we may ignore the term $\Phi''$ in \eqref{phiy} in the first order of $h$). The subsequent terms are obtained in a manner discussed below after the following two remarks.

First, let us analyze the term $\sum\limits_{l=1}^\infty \frac{y_{j+1}^{(l)} + (-1)^l y_{j-1}^{(l)}}{l!}$ in Eq.~\eqref{phiy}. As $y_j$ are functions of $\Phi$ and $y_j = O(h)$, we conclude that $y'_j = O(h \Phi')$ and $y''_j = O(h (\Phi')^2) = o(\Phi'')$. Therefore, the second and the subsequent derivatives of $y_j$ have no impact on $y_{j,m}$ for several first values of $m$, as long as the order of $h$ less than $\Phi''$ is considered. With that remark, we can only leave $y'_{j+1} - y'_{j-1}$ instead of the sum of derivatives in Eq.~\eqref{phiy}.

Second, taking a sum of equations \eqref{phiy} over $j = 0\ldots n-1$ one has
\begin{equation}
    \label{phi_rhs}
    n \Phi''
		=
		h \sum\limits_j \sin \left(\Phi + \frac{2 \pi j}{n}+ y_j\right) 
		= 
		h \sum\limits_j \sum\limits_{l=1}^{\infty} \frac{y_j^{l}}{l!} \sin^{(l)} 
		\left(\Phi + \frac{2 \pi j}{n} \right).
\end{equation}
This allows us to make useful conclusions about the order of $\Phi''$. If we find after obtaining $y_{j,1}, \ldots, y_{j,m}$ as functions of $\Phi$ that the corresponding terms in the sum $\sum\limits_{l=1}^{\infty} \frac{y_j^{(l)}}{l!} \sin^{(l)} (\Phi + \frac{2 \pi j}{n})$ give zero when summed over $j$, then $\Phi''$ has the order at least $m+2$ in $h$. This means that we do not have to take $\Phi''$ into account in Eq.~\eqref{phiy} when finding the next correction $y_{j, m+1} \sim h^{m+1}$. 

Now we describe the details of applying perturbation theory to Eq.~\eqref{phiy}. It is now evident that it is equivalent within the first several orders of $h$ to 
\begin{equation}
    \label{y_eq}
    y_{j+1} - 2 y_j + y_{j-1} = h \sin \left( \Phi + y_j + \frac{2 \pi j}{n} \right) - y'_{j+1}+y'_{j-1}, \quad j = 0 \ldots n-1.
\end{equation}
When considering the order $h^m$ of these equations, the left hand side is $y_{j+1,m}-2y_{j,m}+y_{j-1,m}$, while only $y_{j, m'}$ with $m'<m$ contribute to the right hand side. We look for the solutions in the form
\begin{equation}
    \label{y_sol}
    y_j = \sum\limits_{m=1}^{\infty} y_{j, m} = \sum_{m=1}^{\infty} \sum_{l=1}^{n-1} h^m c_{m l} \sin l \left( \Phi + \frac{2 \pi j}{n} \right) + \Phi' \sum_{m=1}^{\infty} \sum_{l=1}^{n-1} h^m c'_{m l} \sin l \left( \Phi + \frac{2 \pi j}{n} \right) + o(\Phi''),
\end{equation}
where $c_{m l}, c'_{m l}$ are some coefficients. To find them, one has to compare terms in Eq.~\eqref{y_eq} proportional to $h^m \sin l \left( \Phi + \frac{2 \pi j}{n} \right)$ (with or without $\Phi'$). Let's consider terms without $\Phi'$. Substituting \eqref{y_sol} in \eqref{y_eq}, we get
\begin{eqnarray}
    \label{cml}
    &\sum\limits_{m, l} h^m c_{m l} \cdot 2 (\cos \frac{2 \pi l}{n} -1) \sin l \left( \Phi + \frac{2 \pi j}{n} \right) = \\ &= h \sum\limits_{p=1} \sum\limits_{\substack{m_1, \ldots, m_p,\\ l_1, \ldots, l_p = 1}} \frac{h^{m_1  + \ldots+m_p}}{(m_1 + \ldots+m_p)!} c_{m_1 l_1} \cdot  \ldots \cdot c_{m_p l_p} \sin l_1 \left( \Phi + \frac{2 \pi j}{n} \right) \ldots \sin l_p \left( \Phi + \frac{2 \pi j}{n} \right) \cdot \sin^{(p)} \left( \Phi + \frac{2 \pi j}{n} \right)
\end{eqnarray}
The product of sines in the right sum can be expanded as a combinations of sines of $k \left( \Phi + \frac{2 \pi j}{n} \right)$ for $k \leq l_1 + \ldots + l_p + 1$, with the coefficient before $\sin (l_1+\ldots + l_p+1) \left( \Phi + \frac{2 \pi j}{n} \right)$ being $1/2^p$. Firstly, it allows us to compute the coefficients recursively. Secondly, this means that, if $c_{m' l'} = 0 $ for $l'>m'$ when $m' \leq m$, then the power of $h$ is greater than or equal to the coefficient in the sine in each term of the right hand side up to $h^{m+1}$. Comparing it with LHS, we get that $c_{m+1, l} = 0$ for $l>m+1$. A similar relation takes place for $c'_{m l}$ when we compare terms with $\Phi'$. This observation allows us to restrict ourselves to only finding leading coefficients $c_m \equiv c_{m m}$ for each $m$. The recurrent formula for them is
\begin{equation}
    \label{cmm}
    c_m \cdot 2 \left(\cos \frac{2 \pi m}{n} - 1\right) = \sum_{p=1} \sum\limits_{\substack{m_1, \ldots, m_p, \\ m_1 + \ldots + m_p = m-1}} \frac{c_{m_1} \ldots c_{m_p}}{2^p (m-1)!}, 
\end{equation}
where $m \leq n-1$ and $c_1 = \frac{1}{2(\cos \frac{2 \pi}{n}-1)}$ as follows from $y_{j,1}$ found above.

Now we turn again to the expression for $n \Phi''$ given by Eq.~\eqref{phi_rhs}. Aside from the terms with $\Phi'$, it is the same as the right hand side of Eq.~\eqref{cml}, but summed over $j$. Then we can make use of the fact that $\sum\limits_{j=0}^{n-1} \sin m \left( \Phi + \frac{2 \pi j}{n} \right) = 0$ for $m<n $, and when $m=n$ the sum is equal to $n \sin n \Phi$. Since $c_{m l}=0$ when $l>m$ and the same is true for $c'_{m l}$, the term of the lowest order in $h$ in \eqref{phi_rhs} is the one that is given by leading terms of Eq.~\eqref{y_sol} with $m=l$. All terms with $\Phi'$ in Eq.~\eqref{y_sol} give higher orders of $h$. Finally, we have from Eq.~\eqref{phi_rhs} the sine-Gordon equation
\begin{eqnarray}
    \label{sG_arb}
    n \Phi'' &=& h^n {\cal C} \sin n \Phi + o(h^n),\\
		{\cal C} &=& n \sum\limits_{p=1} \sum\limits_{\substack{m_1, \ldots, m_p, \\ m_1 + \ldots + m_p = n-1}} \frac{c_{m_1} \ldots c_{m_p}}{2^p (n-1)!},\nonumber
\end{eqnarray}
where $c_m, m \leq n-1$ are given by recurrent relations \eqref{cmm}. For $n=2$, Eq.~\eqref{sG_arb} takes form $\Phi'' = -\frac{\tilde h^2}{8} \sin 2 \Phi$, which is equivalent to the equation derived in Ref.\cite{maslov}. For $n=3$, Eq.~\eqref{sG_arb} coincides with Eq.~\eqref{phi2}.

\bibliography{Compbib}

\end{document}